\documentclass[12pt,preprint]{aastex}

\slugcomment{To appear in ApJ}
\begin{document}
\title{Extended Red Emission in High-Galactic Latitude Interstellar Clouds}

\author{Adolf N. Witt \altaffilmark{1},  Steve Mandel \altaffilmark{2}, Paul H. Sell \altaffilmark{3}, Thomas Dixon \altaffilmark{4}, \and Uma P. Vijh \altaffilmark{1}}
\altaffiltext{1}{Ritter Astrophysical Research Center, University of Toledo, Toledo, OH 43606}
\altaffiltext{2}{Hidden Valley Observatory, Soquel, CA 95073}
\altaffiltext{3}{Department of Astronomy, University of Wisconsin, Madison, WI 53706}
\altaffiltext{4}{Institute for Astronomy, University of Hawaii, Honolulu, HI 96822}

\begin{abstract}
Nearby interstellar clouds at high Galactic latitudes are ideal objects in which the interaction of interstellar dust with photons from the well-characterized interstellar radiation field can be studied. Scattering and UV-excited photoluminescence at optical wavelengths as well as thermal emission at mid- and far-infrared wavelengths are observable manifestations of such interactions. Here we report initial results from an optical imaging survey of optically thin high-Galactic-latitude clouds, which is designed to study the surface brightness, structure, and spectral energy distribution of these objects. The primary aim of this paper is to study the extended red emission (ERE) that has been reported at high Galactic latitudes in earlier investigations and which is attributed to ultraviolet-excited photoluminescence of an as yet unidentified component of interstellar dust. We conduct this ongoing survey with remotely operated, fast, short-focal-length (0.5 m) telescopes equipped with absolutely-calibrated CCD cameras yielding a field of view of $2\degr\times3 \degr$. The telescopes are located at New Mexico Skies at 7300 ft altitude near Mayhill, NM. The optical surface brightness of our objects is typically a few percent of the brightness of the dark night sky, implying that the cloud SEDs must be deduced from differential surface brightness photometry in different filter bands. We find strong evidence for dust emission in the form of a broad  ($\gtrsim 1000\ $\AA\ FWHM) ERE band with peak emission near 600 nm wavelength and peak intensity of $\sim 5\times10^{-9} \mathrm{erg\ cm^{-2} s^{-1} \AA^{-1} sr^{-1}}$ in optically-thin clouds. This amounts to about 30 \% of the total optical surface brightness of these clouds, the remainder being consistent with expectations for dust-scattered light. This supports claims for the ubiquitous presence of the ERE carrier throughout the diffuse interstellar medium of the Milky Way Galaxy. We suggest that the ERE carrier is involved in the radiative processing of about 20\% to 30\% of the dust-absorbed UV/optical luminosity of the Milky Way galaxy, with the bulk of this energy being emitted in the near- to mid-infrared spectral regions.

\end{abstract}


\section{Introduction}
Extended red emission (ERE) is a widely-observed interstellar photoluminescence phenomenon in the 500 nm to 900 nm spectral range and is seen in many environments where both dust and ultraviolet (UV) photons are present (see Witt \& Vijh (2004) for a recent review). In reflection nebulae (RN) (Witt et al. 1984; Witt \& Boroson 1990), HII regions (Perrin \& Sivan 1992) and planetary nebulae (PN) (Furton \& Witt 1990, 1992) the sky background is faint compared to the relatively high overall intensities of the ERE and can be readily subtracted. However, given the broadband nature of the ERE profile, a persistent problem in observations of the ERE is the separation of the ERE from the normally much more dominant continuum radiation resulting from scattering by dust in the same spectral range in RN (e.g. Witt \& Boroson 1990) or from atomic recombination emission in PN (e.g. Furton \& Witt 1990, 1992), or both, as in the case of HII regions (e.g. Perrin \& Sivan 1992).

Observations have shown that the ratio of ERE to scattered light intensities is highly variable from object to object as well as within individual objects (Witt \& Boroson 1990; Witt \& Malin 1989). Witt \& Boroson (1990) showed for the case of reflection nebulae that this large variation is predominantly due to the variations in the intensity of the scattered light, while ERE, when present, exhibits more nearly constant intensities by comparison. This is most readily explained, if the ERE results from an isotropic emission process, while the scattered light intensity can vary greatly for a constant density of the illuminating radiation simply as a result of differences in the illumination geometry, combined with the strongly asymmetric, forward-directed scattering phase function of interstellar dust. Therefore, a great advantage arises for ERE detection,  when illumination geometries can be found where the dust-scattered light is largely directed away from the Earth by virtue of the forward-directed phase function of scattering of interstellar dust. Such conditions exist in the Red Rectangle nebula (Cohen et al. 2004; Vijh et al. 2006), where the presence of an optically thick dust torus seen almost edge-on directs most of the dust-scattered light in directions perpendicular to the line of sight to Earth. In retrospect, it is easy to see why ERE was first detected in this object (Cohen et al. 1975; Schmidt, Margon, \& Cohen 1980).

 Similarly favorable geometric conditions  exist in high-latitude interstellar clouds, which are illuminated by the integrated radiation field of the Milky Way Galaxy. The intensity distribution of the Galactic radiation field exhibits a pronounced peak in directions close to the Galactic equator. A high-latitude cloud at Galactic latitude $|b|$ receives its most intense illumination under an angle equal to $|b|$ or greater with respect to the line of sight. Consequently, the forward-directed phase function sends most of the scattered light into directions other than the line of sight to Earth (see Figure 5). Therefore, when present in such clouds, the ERE can appear enhanced compared to the underlying scattered light continuum, because the isotropic emission pattern of the ERE does not discriminate against any particular line of sight and the ERE excitation process responds to incident radiation from all direction equally.

The first tentative identification of ERE in the spectrum of a high-latitude dark nebula was made by Chlewicki \& Laureijs (1987), based on the previously published surface brightness spectrophotometry of LDN 1780 by Mattila (1979).  Subsequently, Guhathakurta \& Tyson (1989) determined the optical surface brightness and colors of small portions of the high-latitude counterparts to the Galactic 100 \micron\ cirrus detected by IRAS (Low et al. 1984). They found these to exhibit colors significantly redder than expected for dust clouds scattering the integrated Galactic radiation field. They attributed this red color excess to ERE. Similar results were reported by Guhathakurta \& Cutri (1994), based on the analysis of extensive optical imaging data of several high-latitude clouds. Gordon et al. (1998) conducted the first large-scale survey ($1135 \sq\degr$) designed to detect ERE at high Galactic latitudes, using photometric all-sky data from Pioneer 10 and Pioneer 11 obtained outside the zodiacal dust belt of the solar system. While providing data for large sections of the high-Galactic latitude sky on the one hand, this survey suffered from the relatively large instantaneous field of view of the Pioneer photometers of $2.29\degr \times 2.29\degr$. As a consequence, faint background stars contributed substantially to the measured signal, requiring deep star counts to correct for this stellar contribution.  Also, the large field of view of the Pioneer photometers made it impossible to image individual high-latitude clouds, which typically have angular scales comparable to or smaller than the Pioneer diaphragm apertures. The observations yielded integrated, star-corrected surface brightness data that were well-correlated with HI column densities from large-scale HI surveys and equally well-correlated with the IRAS 100 \micron\ maps of the Galactic cirrus. The investigation of the optical Galactic background radiation corrected for stellar light contribution resulted in the detection in a strong red excess over color predictions from a radiative transfer model for the high Galactic latitude diffuse Galactic light.  This interpretation was confirmed spectroscopically by Szomoru \& Guhathakurta (1998), who obtained spectra of several high-latitude clouds, which revealed the spectral signature of the ERE and confirmed the ERE intensities deduced by Gordon et al. (1998).

 Recently, Zagury (2006) published a re-analysis of the measurements and analysis of Gordon et al. (1998) and concluded that ERE was not a likely component of the high-latitude diffuse Galactic light.  The resolution of this controversy is of importance, because so far only the high-latitude observations of the ERE have provided the basis for an estimate of the photon conversion efficiency with which UV photons are being converted into ERE photons as well as an estimate of the relative abundance of the ERE carrier. Both are crucial constraints in the ongoing efforts to identify the carrier of the ERE.

		Here we report first results from an ongoing imaging survey of high Galactic latitude interstellar clouds (HGLIC), which has been designed to overcome most of the shortcomings of earlier efforts to identify the ERE in high-Galactic latitude dust environments. With this work we follow in the footsteps of Struve \& Elvey (1936) and of Mattila (1970), who showed that the surface brightness of individual high-latitude dark nebulae, which mainly results from scattering of the integrated Galactic star light, can indeed be measured, and of Sandage (1976), who showed that the full complexity of the high-latitude Galactic cirrus can be captured through large-field optical imaging with fast ground-based telescopes. Our survey objectives included the following desirable characteristics: (1) a large field of view to permit the simultaneous imaging of entire clouds and the surrounding sky, making possible reliable background subtraction; (2) sufficient angular resolution to allow nebular surface brightness measurements without contamination from faint stars; (3) use of four well-placed filter bands to define the approximate shape of the spectral energy distribution (SED) of the clouds; (4) focus on optically thin clouds to assure a simplified radiative transfer analysis; (5) a sufficiently large sample of clouds for representative results; (6) fast telescope optics to assure high diffuse-source sensitivity; and (7) reliable absolute intensity calibration. The successful implementation of these essential factors into a pilot observing program will be described in \S 2 of this paper. 

		We describe the data reduction and the resulting SEDs of our objects in \S 3 of this paper. This is followed in \S 4 with an analysis that includes radiative transfer considerations leading to predictions for the expected cloud SEDs on the assumption that scattering by dust of the integrated Galactic radiation field is the sole source of the clouds' optical surface brightness. We will derive red excesses which we interpret as ERE. In \S 5 we will discuss the larger significance of our result for the interstellar medium in general, followed by our conclusions in \S 6.

\section{Observations}

\subsection{Instrumentation}
The imaging observations for this paper were carried out with a Takahashi FSQ 106  f/5 refractor of 106 mm aperture, operated remotely at New Mexico Skies Observatories (NMS) near Mayhill, NM, at an altitude of 7300 feet. The telescope was equipped with a Santa Barbara Imaging Group STL-6303E CCD camera, covering a field of view of $2\degr\times3\degr$. The angular resolution determined by the 9 \micron\ pixels was 3.60\arcsec\ pixel$^{-1}$. The system was equipped with both internal and external automatic guiders and was capable of automatic self-focusing between sub-exposures, a procedure followed routinely. The 50 - mm filters, consisting of four Astro-Don broadband and one Custom Scientific narrow-band H-alpha filters, were housed in a 5-position filter wheel. The band pass characteristics of the four broad filters are summarized in Table 1, and the transmission curves are displayed in Figure 1.  Two of the band passes (G, R) were sensitive to the presence of ERE, while the remaining broadband filters (B, I) were selected to define the scattered-light continuum at either end of the ERE band. The narrow-band H-alpha filter was used to assess the contribution of large-scale H-alpha emission to the surface brightness  of HGLICs detected in the R-band. The entire system was mounted on a Software Bisque Paramount ME equatorial mounting and was operated remotely through the internet.

\subsection{Objects}
Five HGLIC were selected for this pilot study. The clouds are identified by their number in the MBM catalog (Magnani et al. 1985, 1996); their Galactic coordinates and LSR radial velocities are listed in Table 2. The membership in the MBM catalog implies that these clouds have been detected by their CO line emission, and the LSR radial velocities are based on CO observations by Magnani et al. (1985). In today's terminology these clouds would be classified as diffuse molecular clouds (Snow \& McCall 2006) or translucent clouds. Our five clouds were selected from the large MBM sample to be optically thin in the B-band. We will show that the clouds met this requirement in \S 3.1.1 below.

Individual distances of HGLIC are generally quite uncertain, but clouds with V$_{LSR} <$ 10 km s$^{-1}$ typically are part of the thin layer of Galactic interstellar medium which defines the Galactic plane. As a result, such clouds have average z-distances of $\sim$75 pc; at latitudes of $\vert b\vert > 25\degr$ the distances from the sun are in the range from 80 pc to 180 pc (Magnani et al. 1985). Individual distances can be estimated by several techniques, among them reddening observations of stars as a function of distance in the line of sight, or observations of interstellar absorption lines. Penprase (1992, 1993) derived a distance to MBM 30 of 110 pc$\pm$10pc while Magnani \& de Vries (1986) estimated the distance to MBM 32 as $<$ 275 pc.  Consequently, these objects are among the closest individual molecular clouds in the Milky Way galaxy. 

MBM 41A and MBM 41D are sections of the Draco nebula. We follow the designations introduced by Herbstmeier et al. (1993) in defining the sub-regions A and D. The Draco nebula with its components MBM 41A and MBM 41D is a larger complex of intermediate-velocity gas which appears to be falling towards the Milky Way plane. Despite several intensive efforts to determine the distance to the Draco cloud (e.g. Goerigk \& Mebold 1986; Gladders et al. 1998; Penprase et al. 2000), its value remains uncertain. It is likely that the Draco cloud distance is larger than those of the other clouds in our sample and might be found in the range derived in the latest attempt by Penprase et al. (2000) of 800 pc to 1300 pc. Even at these values, the distance remains small in comparison to the diameter of the Galactic disk, which implies that the density of the incident Galactic radiation field should still be similar to that experienced by the other clouds in our sample, which in turn should be close to that in the solar vicinity. However, the lack of accurate distances makes the interpretation of star counts (e.g. Goerigk et al. 1983 for the MBM 41 clouds) highly problematical, because they are seen against a star field in which the volume density declines approximately exponentially with distance, as shown by Mebold et al. (1985). This results in a strong degeneracy involving distances and optical depths for these clouds.

\subsection{Data}
The imaging observations were conducted during photometric nights during the months of January through April, 2006. The data were accumulated in the form of individual sub-exposures of 900 second duration. The dates of observations and the total integration times for the images used in this study are listed in Table 3. In addition to the nebular observations, photometric observations of the absolute flux standard SAO 90153 were obtained through the same system to permit absolute calibration of the resulting data. Similarly, extinction stars were observed through a wide range of air masses to permit the derivation of coefficients of atmospheric extinction in our four broad-filter band passes for the NMS site. Figures 2 a - e display the R-band images of our five objects. In each object nebular regions labeled 1S through 10S are identified where the nebular surface brightnesses were measured; similarly, regions labeled 1 through 3 indicate the locations where measurements of the sky background were obtained. The cloud and sky regions were selected to be free of stars as faint as m $\sim$ 21 mag and their integrated counts were measured with $324 \sq\arcsec (5\times5$ pixels) apertures. The sky-subtracted nebular counts were corrected for atmospheric extinction and converted into absolute intensities, using the calibration data obtained from the photometry of the absolute flux standard star SAO 90153.

We also obtained far-infrared IRAS 100 \micron\ maps of our five HGLICs from the NASA/IPAC Infrared Archive (http://irsa.ipac.caltech.edu/applications/DUST/). We identified our nebula and sky positions on these maps and determined the 100 \micron\ intensity in units of MJy~sr$^{-1}$. As has been noted before (de Vries \& Le Poole 1985; Guhathakurta \& Tyson 1989; Paley et al. 1991, Guhathakurta \& Cutri 1994;  Zagury et al. 1999), the detailed morphology of the HGLICs as seen at far-IR wavelengths are closely matched by the observable morphology at optical wavelengths, in particular in the R-band. The reason for this is that most HGLICs are optically thin at optical wavelengths as they are at $100 \micron$, and the intensities in both bands result from dust-related scattering or emission processes. We confirmed the close similarity of appearance of our five clouds in the far-IR and optical bands; as an example, we show a superposition of the IRAS 100 \micron\ isophotes on the G-band image of MBM 32 in Figure 3. The close similarity between the far-IR and optical morphologies is further supported by the linear relationship between the sky-subtracted nebular intensities at optical and far-IR wavelengths, as discussed in \S 3.1.1.

We also established that the fields identified as \emph{sky} are by no means dust-free but exhibit an easily detected residual surface brightness in the far-infrared.  The IRAS  $100 \micron$ map clearly reveals that what appears as isolated clouds in the MBM CO survey are merely density enhancement in an environment characterized by density fluctuations, with a typical enhancement factor of the order of three for a region defined as a \emph{cloud} over the surrounding \emph{sky} field accessible within the $2\degr\times3\degr$ fields of view of our instrumentation. The CO morphologies plus their associated small-scale structure (e.g. Sakamoto 2002) mainly reflect localized volumes where the density and total shielding in the 912 \AA - 1118 \AA wavelength range, which contains CO-destroying photons, are sufficiently high to permit CO to survive (van Dishoeck \& Black 1988). Our photometric measurements are therefore the result of differential photometry, in which we subtract not only a common foreground intensity stemming from atmospheric air glow and zodiacal light but also a substantial component of interstellar dust-scattered light and possible ERE arising in the non-zero dust column density associated with our sky positions. The averaged (over 10 nebular positions), sky-subtracted optical intensities of our five HGLICs and the corresponding average infrared 100 \micron\ intensities for cloud and sky regions are listed in Table 4. In the last two columns of Table 4 we add the net IRAS 100 \micron\ intensities of our clouds and the ratio of the total 100 \micron\ intensities at the cloud and sky positions, providing a measure of the column density enhancement for the cloud over its nearby surroundings in the plane of the sky.

\section{DATA ANALYSIS}
\subsection{Radiative Transfer Considerations}
\subsubsection{Optical Depths}

A basic assumption essential for our subsequent analysis is that our five HGLICs are optically thin in the B-band, and by inference, in the other optical bands employed in this study as well. The averaging of the surface brightness measurements over ten nebular areas (Table 4), which is necessary in order to overcome the noise from the dominant diffuse foreground sources (air glow and zodiacal light), can only be justified, if the clouds are optically thin. The same precondition applies to the use of a single-scattering approximation for the radiative transfer (\S 3.1.2). Finally, if ERE is present in our HGLICs, it is excited most likely in the far-UV spectral region (Witt et. al. 2006). Consequently, the ERE would arise predominantly near the cloud surfaces, where most of the interstellar far-UV radiation is absorbed. Only clouds that are optically thin in the visual spectral range would permit the ERE from the far side to be fully transmitted, thus significantly enhancing the detection probability. 

The determination of optical depths of HGLICs is difficult. Given the uncertain distances of HGLICs, star counts against clouds embedded in a star field with exponentially declining volume density are practically useless (Mebold et al. 1985). HI column densities through HGLICs, coupled with a standard gas-to-dust ratio (Bohlin, Savage, \& Drake 1978), at best can yield a lower limit to the optical depth, because an unknown, possibly substantial, fraction of hydrogen is likely in molecular form in clouds where CO is beginning to be stable (Burgh et al. 2007). The far-IR intensity is generally the best direct measure of the dust column density through HGLICs, provided it can be calibrated reliably (Schlegel et al. 1998; Schnee et al. 2005, 2006).

The relationship between the 100 \micron\ intensities and the optical depths of HGLICs has been studied extensively.  Schlegel et al. (1998) have suggested a linear relation between the IRAS 100 \micron\ intensities and the corresponding E(B-V), which is being used widely to estimate dust opacities at high Galactic latitudes. This relationship is based on the assumption of a constant dust temperature along sight lines. However, as shown by Cambresy et al. (2001), the assumption of a single linear scaling factor is not supported by data for the Polaris molecular cirrus, where they find A$_V$ values based on star counts that are smaller by factors of two to three compared to the values predicted by Schlegel et al. (1998). It is not clear whether this disagreement is a consequence of the known problems with star counts for HGLICs. Even if all HGLICs are exposed to the same interstellar radiation field, one would not expect the relationship between IRAS 100 \micron\ intensities and optical opacities to be linear, because the far-infrared emissivity depends sensitively upon the grain temperature, which is expected to be lower in portions of clouds which exhibit sufficient column densities of dust to produce some degree of self-shielding against more energetic UV irradiation (Bethell et al. 2004). Also, denser, more opaque clouds may experience grain growth, leading to a slightly lower grain emissivity at 100 \micron\ with resulting lower grain temperature in contrast to grains in less dense, more optically thin clouds. This is indeed borne out by more recent observations (e.g. del Burgo et al. 2003; Lethinen et al. 2007).

Boulanger \& Perault (1988) correlated the IRAS 100 \micron\ intensities and the integrated HI column densities over the Galactic polar caps and found an average I(100 \micron)/A$_V$ ratio of 15.9 MJy sr$^{-1}$ mag$^{-1}$ for a standard A$_V$/N$_H$ ratio of $5.3\times10^{-22}$ mag cm$^2$ (Bohlin, Savage, \& Drake 1978). A similarly high ratio of I(100 \micron)/A$_V$ ratio $\sim$ 18.7 MJy sr$^{-1}$ mag$^{-1}$ was found by Moritz et al. (1998), who analyzed the X-ray shadow of the Draco cloud, of which our targets MBM 41A and MBM 41D are parts. The absorption of X-rays by the gas in foreground clouds has the advantage that both atomic as well as molecular gas is being sampled, as was done by Bohlin et al. (1978). The investigations of Boulanger \& Perault (1988) and by Moritz et al. (1998) are subject to the same uncertainty, stemming from the assumption of a constant gas-to-dust ratio throughout the Galactic interstellar medium. In fact, Moritz et al. (1998) suggest that the gas-to-dust ratio of Bohlin et al. (1978) is probably too large by a factor of two for the Draco cloud. This is supported by optical observations of stars seen through the Draco cloud (e.g. Penprase et al. 2000), which lead to higher visual extinctions than implied by the total hydrogen column density. Other studies, which rely on A$_V$ values derived from star counts or reddening observations, also find lower ratios of I(100 \micron)/A$_V$ than deduced by Boulanger \& Perault (1988) and by Moritz et al. (1998). However, for A$_V$ values based on star counts to be reliable, the optical depths of the associated clouds must be relatively high, the distances of the clouds need to be known, and the I(100 \micron) values are also much higher than those measured in our five target HGLICs. For example, Boulanger et al. (1998) studied the Chamaeleon molecular clouds, where A$_V$ values up to 2.5 mag are typical, associated with infrared intensities of up to 17 MJy sr$^{-1}$. Depending on the optical depths of individual cloud portions, Boulanger et al. (1998)  found ratios of I(100 \micron)/A$_V$ from $\sim$5 MJy sr$^{-1}$ mag$^{-1}$ for the densest clumps to $\sim$10.5 MJy sr$^{-1}$ mag$^{-1}$ in more transparent regions. Similar values are found by de Vries \& De Poole (1985) and by Zagury et al. (1999) from the analysis of individual high-latitude clouds. The lack of agreement between these results fully supports recent conclusions of Schnee et al. (2005, 2006), who warn about the inherent uncertainties in deriving dust column densities from IRAS 100 \micron\ surface  brightnesses. 
We conclude from these studies that for optically thin clouds as represented by our HGLIC sample a ratio of I(100 \micron)/A$_V\sim$ 10 MJy sr$^{-1}$ mag$^{-1}$ is a fair, albeit uncertain estimate. This conclusion is also supported by the recent investigation of LDN 1642 by Lethinen et al. (2007). 

Fortunately, for the purpose of ERE detection, \textit{the precise values of the optical depths of our HGLICs need not be known, provided they are entirely in the optically thin regime.} We are able to establish this empirically by plotting measured B-band net surface brightnesses of HGLICs against the corresponding IRAS 100 \micron\ intensities at identical positions. We are presenting such a plot in Figure 4 for MBM30, MBM32, MBM41A, and MBM41D, where the IRAS 100 \micron\ data are measured relative to a sky reference level of 0.65 MJy sr$^{-1}$. The relationship is linear and these four clouds obey the same relationship. Since these clouds are optically thin at 100 \micron, this shows that they are also optically thin in the B-band, and that these clouds are exposed to the same effective radiation field. De Vries and Poole (1985) have shown that the B-surface brightness of HGLICs begins to saturate for A$_V >0.7$ mag. This implies that the optically thin regime extends to about A$_V\sim$ 0.7 mag, or $\tau_B\sim$ 0.83. Figure 4 shows that our clouds are well within this linear, optically-thin regime. This is reflected as well by the average optical depths determined for the individual clouds in Table 5. After converting the optical intensity units shown in Figure 4 into MJy sr$^{-1}$ (Leinert et al. 1998), we find

\begin{equation}
	I_B \left[MJy\ sr^{-1}\right]  =  (2.13\pm0.10)\times 10^{-3} (I_{100} - 0.65) \left[MJy\ sr^{-1}\right] . 
\end{equation}

The slope in Eq. 1 agrees closely with a similar value found by Zagury e al. (1999), who derived a slope of $(2.4\pm 0.2)\times 10^{-3}$ for the cloud MCLD123+24.9, although those authors attributed the illumination of that cloud to the star Polaris. MBM 25 does not follow this relationship (Eq. 1) but rather a linear relation with a slope approximately twice as steep. This indicates that MBM 25 is also optically thin but is illuminated by a radiation field which is either more intense or is scattered more effectively toward Earth. To avoid confusion, however, the data for MBM 25 are not included in Figure 4.

We estimate average values of the net extinction of A$_V \sim 0.4$ mag deduced from the (cloud - sky) 100 \micron\ intensities found for our five HGLICs in Table 4, with a spread of a factor two. This is supported in general by the study of Larson \& Whittet (2005) who determined the extinction to over 100 stars located behind high-latitude translucent clouds. The peak in their distribution of extinctions is found at A$_V <$ 0.5 mag. Thus, we conclude that the radiative transfer in our objects occurs under optically thin conditions and that multiple scattering and significant self-absorption are not important in our objects. 

\subsubsection{Expected Scattered Light Intensities}

With ERE detected in other environments at wavelengths longer than 500 nm, only our measured intensities in the B-band can be assumed to be dominated by dust-scattered light, generated through external illumination by the Galactic interstellar radiation field. The subsequent section will therefore be restricted to estimating the expected surface brightness of our HGLICs in the B-band via a number of approximations with increasing complexity.

A critical component of any model prediction of HGLIC intensities is the density and directional distribution of the Galactic interstellar radiation field (IRF) in the solar vicinity. We used the recent determination of the IRF by Porter and Strong (Porter \& Strong 2005; Moskalenko et al. 2006). We are grateful to Troy Porter for providing us with data regarding the directional distribution of the intensity of the IRF at the galactocentric radius of 8.5 kpc, both for a location in the plane (z = 0 pc) and for locations appropriate for many HGLICs, i.e. z = 100 pc. 
In Figure 5 we show an Aithoff projection of the radiation field for z = 100 pc for the wavelength of our B-filter, centered on the direction to the Galactic center. The positions of the HGLICs included in this study are indicated on this map, with MBM 41A representing both MBM41 regions. Porter \& Strong find a radiation density of U$_\lambda = 5.86 \times 10^{-17} \mathrm{erg\ cm^{-3}\ \AA^{-1}}$, corresponding to an average intensity of J$_\lambda = 1.4 \times 10^{-7} \mathrm{erg\ cm^{-2}\ s^{-1}\ \AA^{-1}\ sr^{-1}}$ for the B band. These values are in near perfect agreement with the earlier determination of the IRF  density in the solar vicinity by Witt \& Johnson (1973) and are only about 16\% higher than the predictions for the average intensity of the IRF by Mathis et al. (1983). The latter IRS intensities were known to be too low to fully account for the observed far-infrared emission from the Milky Way galaxy (Mathis 1998; Gordon et al. 1998), a deficiency corrected in the recent self-consistent calculation of the IRF by Porter and Strong.

Early studies of the surface brightness of HGLICs (Sandage 1976; Guhathakurta \& Tyson 1989) employed a single-scattering approximation in the optically-thin limit, assuming isotropically scattering grains with an albedo $a$ = 0.6, to estimate the expected surface brightness  $I_B$ of HGLICs. Thus, 
\begin{equation}
	I_B \approx \frac{c}{4\pi}u_B  a  \tau_0     
\end{equation}

\noindent where $\tau_0$ is the line-of-sight optical depth through a cloud at the effective wavelength of our B-filter. Assuming $\tau_0$ = 0.5 and the radiation density $u_B = 5.86 \times 10^{-17} \mathrm{erg\ cm^{-3}\ \AA^{-1}}$ from Porter and Strong, this approximation yields $I_B = 42\times10^{-9} \mathrm{erg\ cm^{-2}\ s^{-1}\ \AA^{-1}\ sr^{-1}}$.  A comparison with our results in Table 4 shows that this estimate exceeds the intensities actually observed in the B-band by factors of two to three. While some of our clouds have values of $\tau_0 < 0.5$, we attribute the higher-than-observed intensity primarily to the failure of the isotropic scattering assumption, given that it is well established that the phase function of interstellar grains is strongly forward directed (Gordon 2004). In a high-latitude cloud, an isotropic phase function is unable to respond correctly to the highly non-isotropic IRF in which the illumination comes primarily from a ring-shaped source following approximately the Galactic equator. The flux from this principal source is incident on a HGLIC at Galactic latitude $|b|$ under scattering angles of $\geq b$, and is thus scattered mostly not toward the Earth. A second contributing factor results from the fact that $\tau_0$ = 0.5 is already approaching the limits of the optically thin regime, which ignores all effects of self-shielding. 

The latter limitation can be overcome by employing the exact radiative transfer solution for isotropic scattering for an externally illuminated cloud provided by van de Hulst (1987), which takes self-shielding and multiple scattering into account. His Fig. 3 presents the Bond albedo A for spherical clouds over a wide range of optical depths and grain albedos $a$. For an optical diameter = 0.5 and a dust albedo $a$ = 0.6, we interpolate to find $A$ = 0.18. The definition of the Bond albedo employed by van de Hulst then leads to

\begin{equation}
	I_B = A\  J_B
\end{equation}

where $J_B$ is the average intensity of the IRF, averaged over the entire sky. With $J_B$ taken from Porter and Strong, we find $I_B = 25 \times 10^{-9} \mathrm{erg\ cm^{-2}\ s^{-1}\ \AA^{-1}\ sr^{-1}}$. This is much closer to the observed intensities but still somewhat higher than the average of the observed values. The actual, forward-directed scattering phase function of interstellar dust grains responds most directly to the line-of-sight intensity of the IRF, which is less than the average $ J_B$ for HGLICs observed from a location near the Galactic plane. This may explain the remaining difference between observed and predicted intensities.

Witt (1985) published a simple single-scattering approximation for the scattered intensity including self-shielding, which should work equally well, provided we remain near the optically thin limit. For an externally illuminated cloud immersed in a radiation field with average intensity $J_B$ we find from Witt (1985)

\begin{equation}
	I_B \approx  (1 - e^{-\tau_0} )  e^{-\tau_{abs}}  a  J_B	
\end{equation}

where $\tau_{abs}$ is the average absorption (in contrast to extinction) optical depth experienced by incident photons before their first scattering. For isotropically incident radiation we approximate $\tau_{abs} \simeq 0.5 (1-a) \tau_0$. With the same parameters as used in the previous examples we find $I_B  \simeq 27 \times 10^{-9} \mathrm{erg\ cm^{-2}\ s^{-1}\  \AA^{-1}\ sr^{-1}}$, very similar to the result obtained from the exact solution by van de Hulst (1987) for the same case, as well as relatively close to the observed intensities. 

Eq. (4), as was the case with the other approximations, fails to take into account the forward scattering nature of interstellar grains. This weakness was overcome in the approximation presented by Stark (1995), who treated the transfer of radiation through an externally illuminated HGLIC of $\tau_0 =1$ with grains characterized by $a$ = 0.6 and a phase function asymmetry parameter $g$ = 0.75, representing a strongly forward-directed phase function. The most severely limiting approximation in this case was Stark's assumption that the illumination source containing the IRF was confined to a geometric sheet located below the cloud under investigation. Incident radiation is thus limited in origin to a hemisphere, with an average intensity of $2\ J_B$. As expected, Stark finds a strong dependence on Galactic latitude in the amount of light back-scattered to the observer. Interpolating his results for $\tau_0  =  0.5$ for $\vert b\vert = 30\degr$, the reflected fraction of the average incident intensity is 0.033. Thus, $I_B  \simeq  9.2 \times 10^{-9} \mathrm{erg\ cm^{-2}\ s^{-1}\ \AA^{-1}\ sr^{-1}}$. This value represents only  about 50 \% of the average B-intensities observed in our HGLICs.  This under-prediction of the intensity most likely is caused by the fact that the model ignores the non-zero illumination originating with the IRF produced by sources in the upper hemisphere above the cloud.  A comparison of the simple models based on single scattering with an isotropic phase function with the model of Stark, which includes forward scattering but assumes an unrealistic source distribution, suggests that the former produce better results, as long as the HGLICs considered are located at intermediate Galactic latitudes, as is the case with our objects, and as long as self-shielding is included, as in Eq. (4).

\subsubsection{Scattered Light SEDs of HGLICs}
The observed optical SEDs of HGLICs are expected to consist of two components, the scattered light continuum produced by scattering of the interstellar radiation field by the dust concentrated in the HGLICs, and the ERE which is expected to be observable at wavelengths longer than 500 nm, if ERE is present. The presence of ERE can be established by the comparison of the observed SEDs with an expected scattered light SED, yielding an excess in the red wavelength region with the usual spectral characteristics of the ERE. Based on the considerations of Sect. 3.1.2, we have adopted the approach of Witt (1985), as expressed in Eq. (4), to predict the expected scattered light SEDs of HGLICs. Two specific adjustments were made, however. First, to reflect the fact that our surface brightness measurements represent differential photometry measurements between nebular fields and background ``sky'' fields, which themselves have dust column densities roughly half of the corresponding nebular positions, we calculated the anticipated scattered light intensities for both sky and nebular fields separately and used the differences for comparison with our observed intensities. Second, we normalized the predicted scattered light SEDs at 454 nm to bring models and observations in agreement at a wavelength where the observed intensity is expected to consist of scattered light alone. This normalization corrects for the fact that the scattering model does not include a forward-directed scattering phase function and also does not include the true directional distribution of the illuminating interstellar radiation field but uses the average intensity instead. This normalization is justified by the facts that both the phase function asymmetry of the scattering dust grains (Draine 2003) and the geometry of the interstellar radiation field (Porter \& Strong 2005) are varying only slightly over the wavelength range covered by our observations. On the other hand, our normalized models fully preserve the much more pronounced wavelength dependencies of the dust optical depth, the grain albedo, and the average intensity of the illuminating radiation field (see Table 6).

Accordingly, scattered light SEDs were calculated from the expression
\begin{equation}
\Delta I_\lambda  \approx a_\lambda   J_\lambda   \left\lbrace  (1 - e^{-\tau_0^\lambda})  e^{-\tau_{0,abs}^\lambda} - (1 - e^{-\tau_{sky}^\lambda})  e^{-\tau_{sky,abs}^\lambda}\right\rbrace 
\end{equation} 
where $a_\lambda$ is the albedo for interstellar dust with R$_V$ = 3.1 from Draine (2003) and $J_\lambda$ is the average intensity of the Galactic interstellar radiation field in the solar vicinity (R = 8.5 kpc, z = 100 pc) obtained from Porter \& Strong (2005). Both quantities are listed for 15 wavelengths covering the spectral range of our filters in Table 6. The optical depths  $\tau_0^\lambda$ and $\tau_{sky}^\lambda$ are the line-of-sight optical depths through the HGLICs and their reference sky fields, respectively; they are estimated from the corresponding 100 \micron\ surface brightness found in the IRAS all-sky survey by using the relation $\tau_{454} = 0.12 I_{100 \micron} [MJy sr]^{-1}$ for optically thin clouds, according to our discussion in Sect. 3.1.1. The optical depths $\tau_0$ and $\tau_{sky}$ are scaled as a function of wavelength according to the law of interstellar extinction for R$_V$ = 3.1, as given in Table 6. As in Eq. (4), we approximate the self-shielding absorption optical depth $\tau_{0,abs}  = 0.5 (1-a) \tau_0$ and $\tau_{sky,abs} = 0.5 (1-a) \tau_{sky}$. This approximation reflects the fact that self-shielding only involves the non-scattered fraction, i.e. (1 - $a$ ), of the incident light and that in externally illuminated clouds the average optical path for reaching a point of interaction is half the total optical depth through the cloud, assuming perfect symmetry. The self-shielding absorption optical depths are scaled with wavelength by using the optical depth scale and the wavelength-dependent albedo values listed in Table 5.

\subsection{ERE intensities}
The comparison of the observed surface brightness values and the predicted cloud SEDs based on scattering alone (Figures 6a - e ) shows that all five HGLICs exhibit a significant excess in the R filter, which is consistent with the presence of ERE. The ERE spectra of the different clouds in our sample, which are obtained by subtracting the scattered-light SED from the observed surface brightnesses, show interesting differences, however. In three cases (MBM25, MBM41A and MBM41D), the ERE intensity in the G- and R-bands are nearly the same, suggesting a peak in the ERE spectrum near 600 nm. In one cloud, MBM 32, the excess is limited to the R-band alone, while in MBM 30 the highest ERE intensity occurs in the G-band, followed by a lower intensity in the R-band. These differences appear to be real, and they are consistent with the well-established fact that the peak of the ERE band occurs at different wavelengths in different objects (Witt \& Boroson 1990; Smith \& Witt 2002). Further support for this conclusion can be found in the near-identity of the ERE spectra for MBM 41A and MBM 41D. The two data sets of these two separate parts of the Draco cloud were obtained at different times and were treated as independent data throughout the analysis; yet, they suggest that the Draco cloud has an ERE spectrum that does not vary within the cloud. The observed intensities in the I-band essentially reverts to the predicted level for scattering alone for four clouds. Only MBM 25 exhibits a significant excess in the I-band as well. This is not inconsistent with ERE seen in other sources; on the contrary, I-excesses in ERE-bright regions are very common in reflection nebulae (Witt \& Schild 1985). The BGRI filter system employed in the present study provides insufficient resolution for a detailed investigation of differences in the ERE spectra of HGLICs. 

The scattered-light subtracted ERE intensities in our five HGLICs are remarkably constant, typically $I_{ERE} = (5\pm1) \times 10^{-9}\mathrm{erg\ cm^{-2}\ s^{-1}\ \AA^{-1}\ sr^{-1}}$. The width of the ERE band is difficult to estimate with only four filters, but the data in Figures 6a - e suggest FWHM $\gtrsim 1000 \AA$. In contrast to the near-constancy of the ERE intensity, the scattered light intensities within our sample of five HGLICs vary by almost a factor of two. This range of scattered-light intensities matches almost exactly the range of the quantity (cloud - sky) in Table 4, the excess 100 \micron\ intensity of the five HGLICs over that of the respective surrounding sky, which is a direct measure of the cloud optical depths. However, the correlation between the optical depths of the individual clouds and the scattered light intensity is not perfect, e.g. the cloud with the brightest scattered light intensity, MBM 25, has the second lowest optical depth. Thus, we conclude that the observed range of scattered light intensities must also be influenced by differences in the directional distribution of the illuminating radiation field. This should be expected, given the wide range of Galactic longitudes of the HGLCIs in our sample (see Table 2). We note, therefore, that the results of our decomposition of the observed intensities into a relatively variable scattered scattered light component and a comparatively constant ERE component is consistent with our original expectations. The near-constancy of the ERE is primarily caused by an isotropic emission process, which is responsive to the average intensity of the incident radiation field while remaining insensitive to its directional distribution.  The scattered light intensity, on the other hand,  is strongly coupled to the directional anisotropy of the incident radiation through the forward-directed scattering phase function of the interstellar grains.

A second factor is likely to contribute to the different degrees of variability of the intensities of ERE and scattered light in the case of clouds with a range of optical depths. It is known that ERE requires far-UV radiation for its initiation (Witt \& Schild 1985; Darbon et al. 1999; Witt et al. 2006). The optical depth of a given cloud is between two and three times larger in the wavelength region of ERE excitation compared to the optical depth at the wavelength of our B-filter (457 nm). While our clouds are optically thin at B (Sect. 3.1.1), they will approach the optically thick condition in the far-UV, where ERE is excited. Thus, nearly all UV photons capable of initiating ERE are absorbed and the resulting ERE intensity saturates to a constant level, arising mostly in the surface layers, independent of the optical depth. By contrast, scattering at optical wavelengths, where the clouds are optically thin, occurs throughout the cloud and the resulting scattered light intensities vary with the differences in optical depth.

Finally, we must recognize that the ERE and scattered light intensities obtained from our differential photometry data are always lower limits, because the column density of interstellar dust associated with the "sky" positions generates both forms of radiation, with intensity levels comparable to the values shown in Figures 6a - e. These sky-related ERE and scattered light intensities are subtracted in the data reduction process. Therefore, to correct for this effect, we estimate that the absolute ERE intensity in HGLICs is probably of the order of $\sim 10\times10^{-9} \mathrm{erg\ cm^{-2}\ s^{-1}\ \AA^{-1}\ sr^{-1}}$. This estimate takes into account that  the ERE intensity is expected to increase linearly at very small optical depth before beginning to saturate at the optical depths found in our "cloud" positions. The ratio of ERE to scattered light intensity is expected to decrease with optical depth. Our HGLIC of smallest optical depth, MBM 41D, has a value of $\sim$ 1/2 for this ratio, which implies that the ERE intensity is approximately 1/3 of the total interstellar radiation in the R-band coming from the low-density diffuse cirrus medium at high latitudes. This agrees remarkably well with the findings of Gordon et al. (1998, Fig. 23) and Szomoru \& Guhathakurta (1998). Also, our estimated total intensity of the ERE, when integrated over the FWHM of the R filter of Pioneer 10/11 (968 A) results in a band-integrated ERE intensity of $9.7\times10^{-6} \mathrm{erg\ cm^{-2}\ s^{-1}\ \AA^{-1}\ sr^{-1}}$, in excellent agreement with the results obtained with Pioneer 10/11 by Gordon et al. (1998).

\section{Discussion}
From the outset, the clouds in our HGLIC sample have been selected to be optically thin.  Every cloud in our sample, albeit limited, exhibits ERE with very similar intensities, apparently unrelated to the optical depths of the clouds, which vary over a range of about a factor two. As a result, the highest ratio of ERE to scattered light intensity occurs in the cloud with the smallest optical depth, MBM 41D. The data fully support the suggestion that the  ERE results from an isotropic emission process that is initiated by the far-UV radiation component of the Galactic interstellar radiation field (Witt \& Schild 1985; Witt et al. 2006). In the case at hand of isolated clouds immersed into the interstellar radiation field, the maximum ERE intensity will be reached quickly when the physical penetration depth for the exciting UV photons begins to equal or becomes less than the average cloud diameter, while the cloud remains still quite optically thin for the resulting ERE photons. This effect is expected to be particularly pronounced at high Galactic latitudes where the wavelength dependence of interstellar extinction is characterized by R$_V$ values smaller than the typical value of R$_V$ = (3.05 $\pm$ 0.15) found for the diffuse ISM in the Galactic plane (Larson \& Whittet 2005). A lower value of R$_V$ generally indicates an extinction curve of increased steepness in the far-UV (Cardelli et al. 1989), which implies a larger than typical ratio of optical depths for specific UV wavelengths to those at optical wavelengths. As a consequence, clouds with a range of optical thicknesses as represented in our sample, while more or less optically thin in the visual range, are close to being optically thick in the far-UV, with a resultant saturation in the ERE excitation.

The near-constancy of the observed ERE intensities has further important implications. Given the proximity to Earth of the clouds in our HGLIC sample, we can assume that the average intensity and spectral energy distribution of the illuminating interstellar radiation field is essentially the same for all clouds and similar to the interstellar radiation field in the solar vicinity. Then, our results imply that the abundance of the ERE carrier compared to the abundance of dust responsible for the UV/optical interstellar extinction is relatively constant. The ERE carrier must, therefore, be a common constituent of the interstellar medium in diffuse interstellar clouds, which makes a significant contribution to interstellar extinction through absorption, as we will show below.

Gordon et al. (1998) and Szomoru \& Guhathakurta (1998) compared the number of ERE photons emitted to the number of photons absorbed from the interstellar radiation field by the diffuse ISM over the 91.2 nm to 550 nm wavelength range, finding a ratio of 0.10 $\pm$ 0.03. Assuming that at most one ERE photon can be generated by photoluminescence for every UV/optical photon absorbed, they interpreted this ratio as a lower limit to the photon conversion efficiency of the ERE process. Inversely, if we assume that the photon conversion efficiency is at most 100\%, this ratio can also be interpreted as the lower limit to the absorption cross section provided by the ERE carrier as a fraction of the total continuous UV/optical interstellar absorption cross section exhibited by all interstellar dust components. A more realistic estimate for the ERE photon conversion efficiency may be 30\% to 50\% instead of the theoretical upper limit of 100\%. In that case, the ERE carrier contributes from 20\% to 30\% of the total continuous UV/optical absorption cross section of the diffuse Galactic ISM. Depending on precisely where in the UV the absorption by the ERE carrier peaks, only a small fraction of the energy of a photon absorbed by the ERE carrier is re-emitted in the form of optical ERE photons, which have typical energies near 1.8 eV. The remainder of the energy, which could be as much as $\sim$25\% of the total absorbed interstellar radiation field energy, will have to be re-emitted at longer wavelengths, most likely in the near- and mid-IR. This amount is remarkably close to the observed fraction of the total stellar radiation of the Galaxy reprocessed by the ISM and emitted in the near/mid-IR (Boulanger \& Perault 1988). We suggest, therefore, that the ERE carrier is a major contributor to the near/mid-IR emission from the Galactic cirrus, or inversely, that one characteristic of a large fraction of the particles giving rise to the near/mid-IR Galactic emission is the ability to re-emit a small fraction of the energy absorbed in the UV/optical range in the form of optical photoluminescence photons appearing as ERE. It appears likely, therefore, that the ERE is a significant component of the processing of Galactic stellar radiation by interstellar dust. In the controversy concerning the existence of ERE in the diffuse ISM that has arisen between Guhathakurta \& Tyson (1989), Gordon et al. (1998), and Szomoru \& Guhathakurta (1998) on the one hand and Zagury (2006) on the other, our findings clearly support the former group of investigations.

\section{Conclusions}
We have conducted differential surface brightness photometry in four broad optical bands (B, G, R, I) of five optically thin HGLICs, employing the method of wide-field CCD imaging. We compared the resulting intensities with spectral energy distributions expected for scattered light only, assuming that the source of illumination is the Galactic interstellar radiation field of the solar vicinity. We reached the following conclusions:
\begin{enumerate}
\item We detected excess intensities in the R-band in all cases, in the G-band in all but one case, and in the I-band in one case. These excess intensities and their spectral energy distributions are consistent with the presence of ERE in all HGLICs examined in this study. The ERE intensities are remarkably constant for clouds with a range of optical depths of about a factor two.

\item The constancy of the ERE intensities is consistent with excitation of the ERE through far-UV photons, which saturates well before clouds become optically thick in the red region where ERE photons are emitted, and with a near-constant abundance ratio of ERE carrier particles to dust grain causing the interstellar reddening.

\item The absolute ERE intensities in HGLICs, determined by correcting the results of our differential photometry for the ERE contribution expected to be present in our reference sky positions are in excellent agreement with measurements of the ERE in the high-latitude diffuse Galactic light using Pioneer 10/11 photometry by Gordon et al. (1998) and the results of differential spectro-photometry of individual HGLICs by Szomoru \& Guhathakurta (1998).

\item We find that the ratio of  the number of ERE photons and the number of absorbed UV/optical photons in the interstellar radiation field, which leads to an estimate for a lower limit for the ERE photon conversion efficiency of $\sim10\pm 3$\% (Gordon et al. 1998; Szomoru \& Guhathakurta 1998), can also be interpreted to suggest that the ERE carrier particles contribute about 20\% to 30\% of the continuous UV/optical absorption cross section in the diffuse ISM and very likely are responsible for a major fraction of the near/mid-IR radiation emitted by the Galactic cirrus.
\end{enumerate}

\acknowledgments
This research has been supported through grants AST0307307 and AST0606756 from the National Science Foundation to the University of Toledo. Our observing program at New Mexico Skies received additional support from corporate sponsors AstroDon, RC Optical Systems, Santa Barbara Instrument Group, and Software Bisque, for which we are very grateful.

\appendix

\clearpage

\begin{figure}
\includegraphics[height=8.5in]{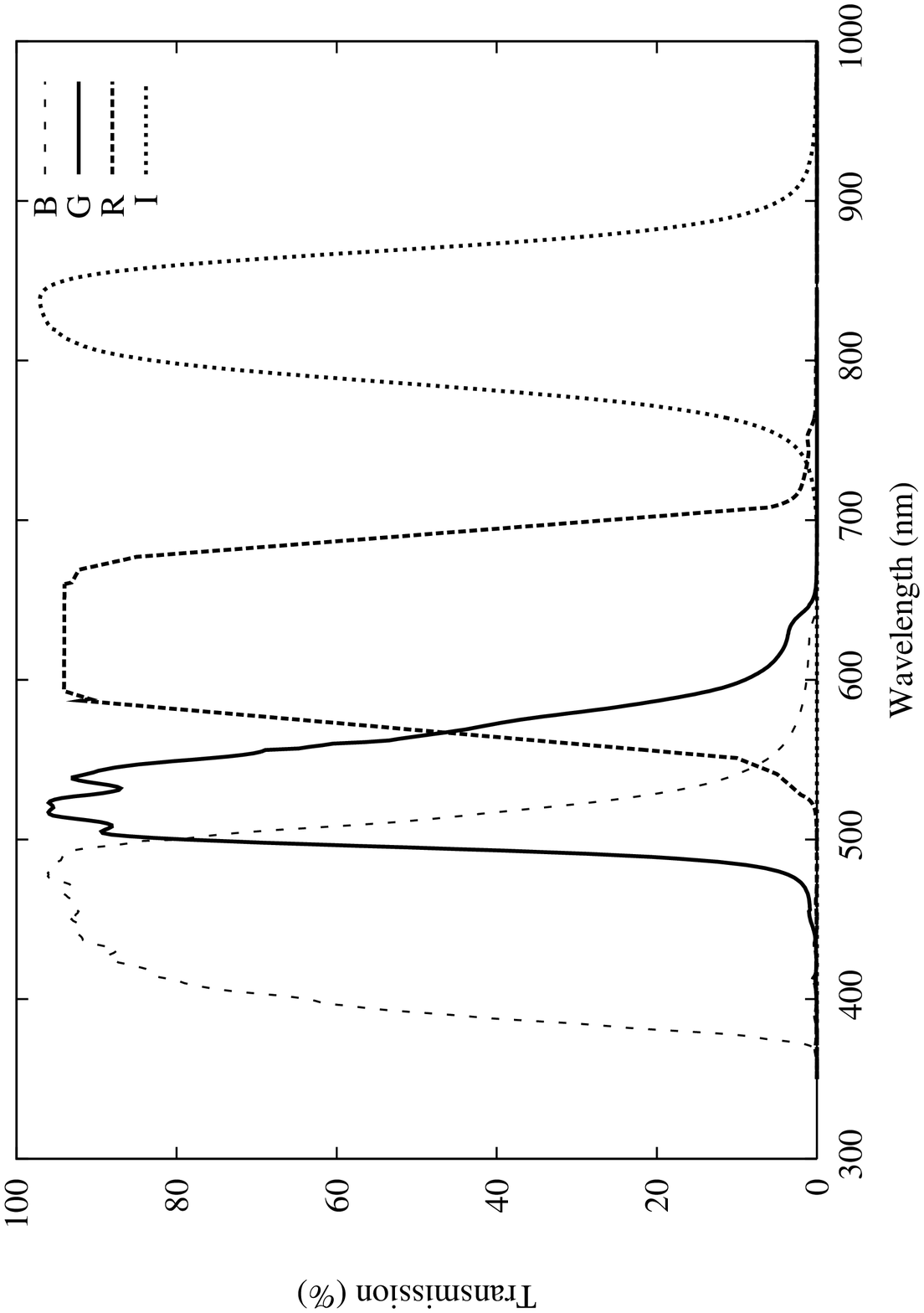}
\caption{Transmission curves for four broad-band AstroDon filters used in this study.}
\end{figure}


\begin{figure}
\centering
\includegraphics[width=6.5in]{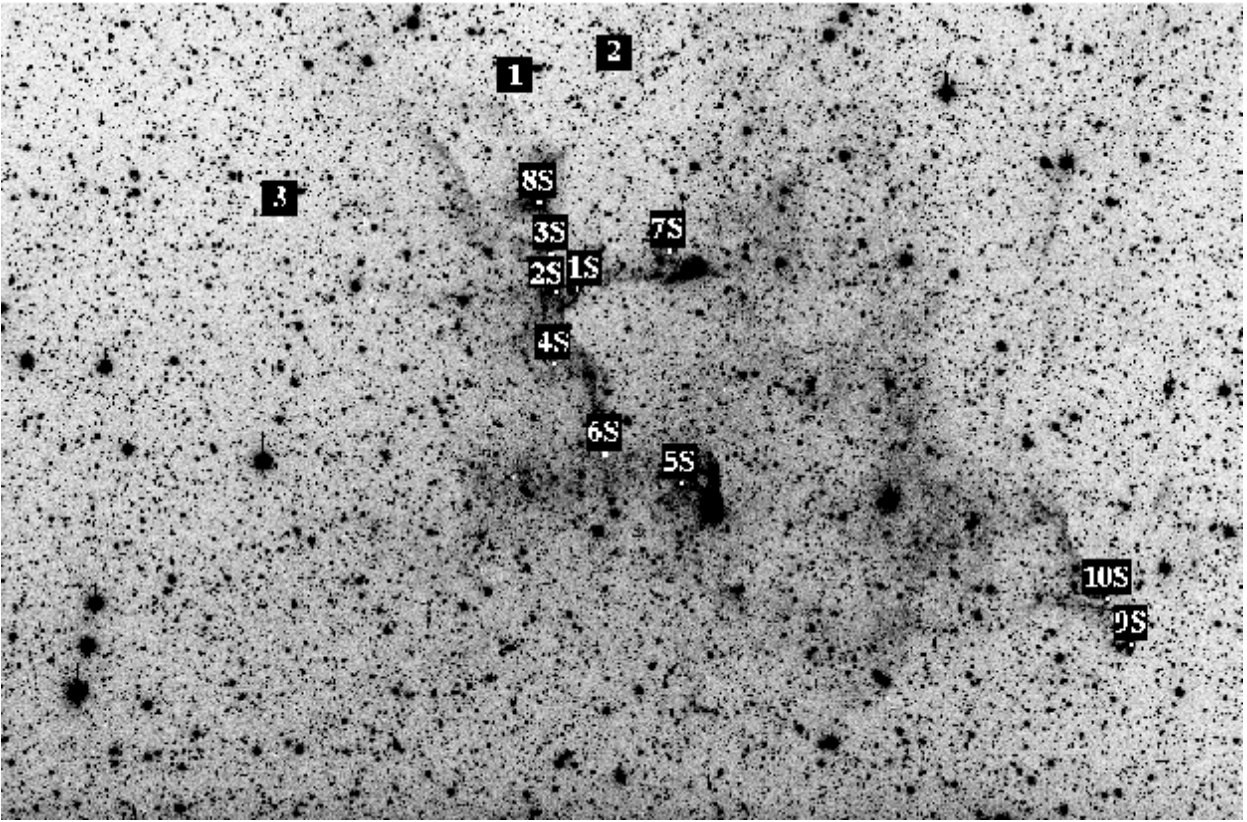}
\caption{(a - e)We display the R-band images of (a) MBM 25, (b) MBM30, (c) MBM32, (d) MBM41A, and (e) MBM41D. The image sizes are $2\degr\times3\degr$, with North up and East to the left. Reference sky positions are indicated by labels 1, 2, 3 while the cloud positions at which surface brightness measurements were conducted are labeled 1S trough 10S in each case.}
\end{figure}
\centering{
\includegraphics[width=6.5in,clip]{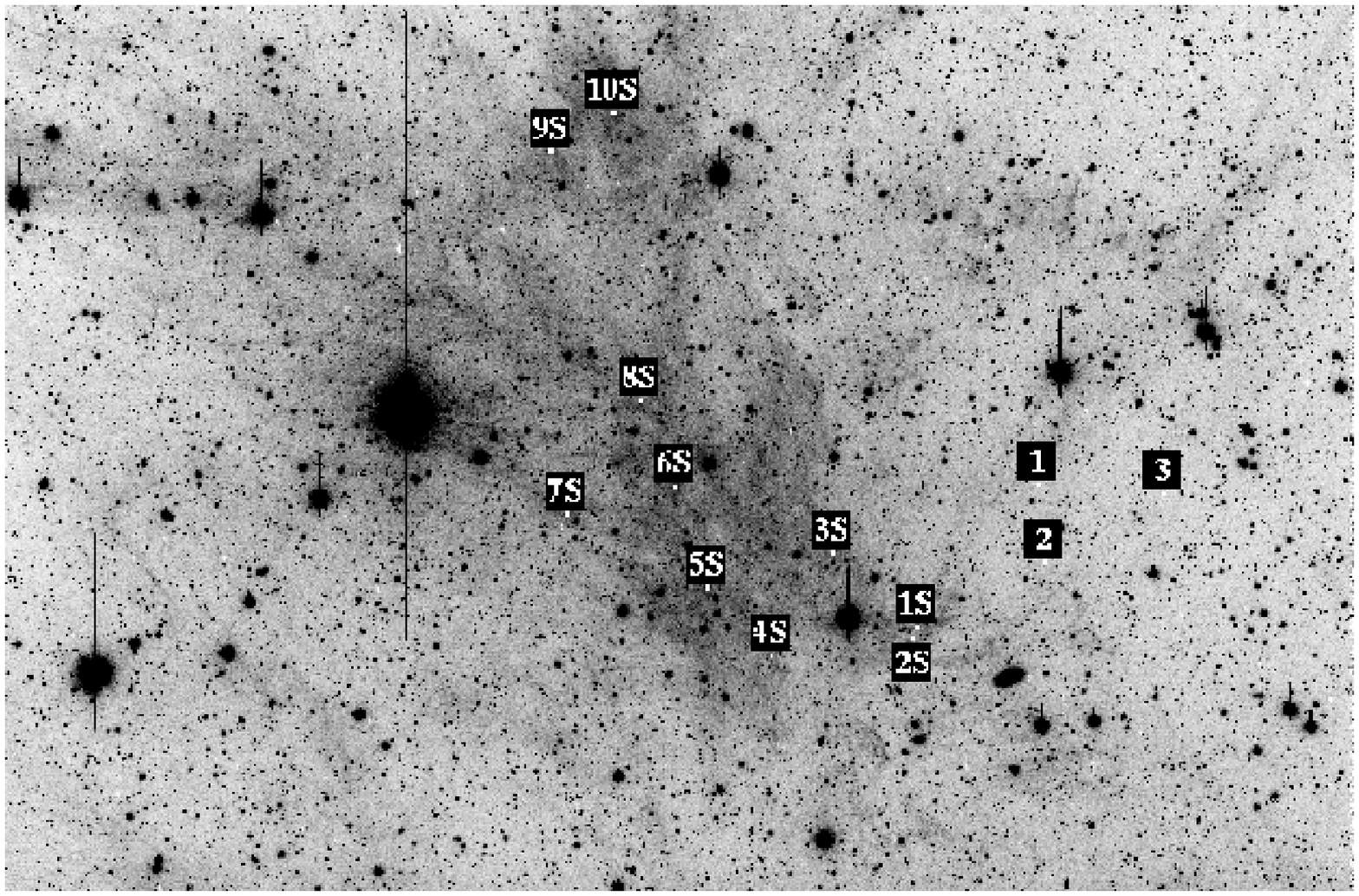}
\includegraphics[width=6.5in,clip]{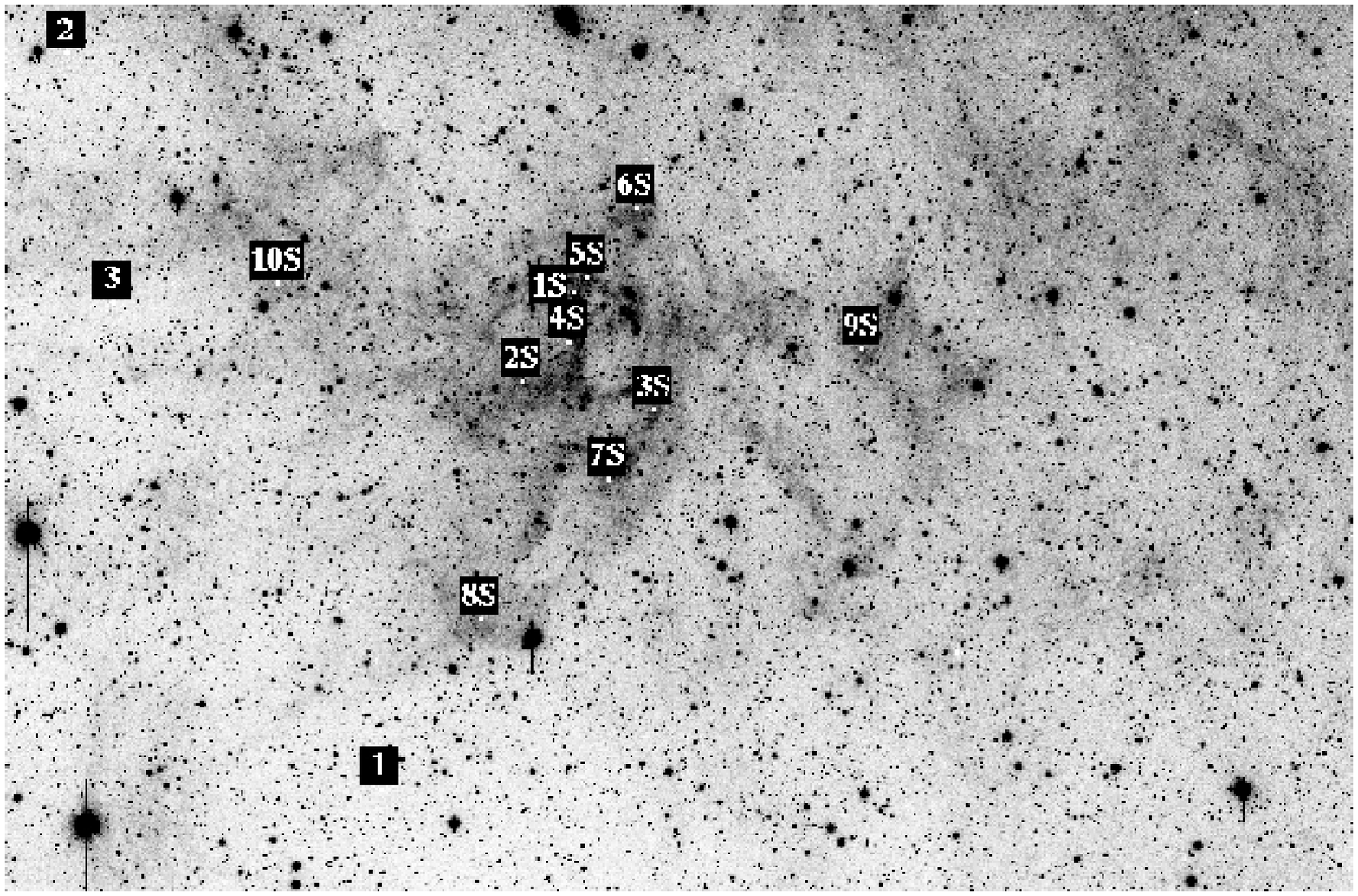}
\includegraphics[width=6.5in,clip]{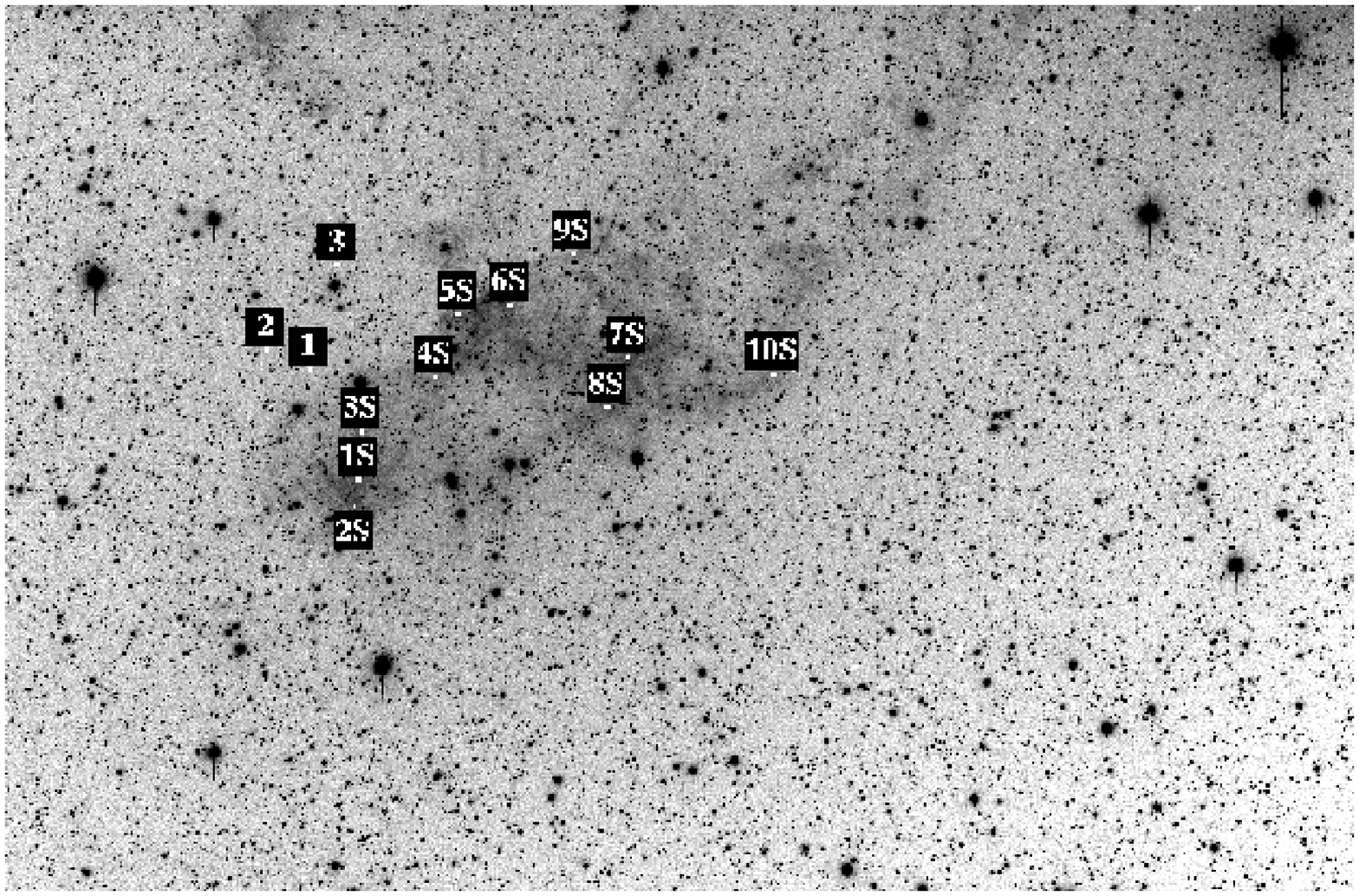}
\includegraphics[width=6.5in,clip]{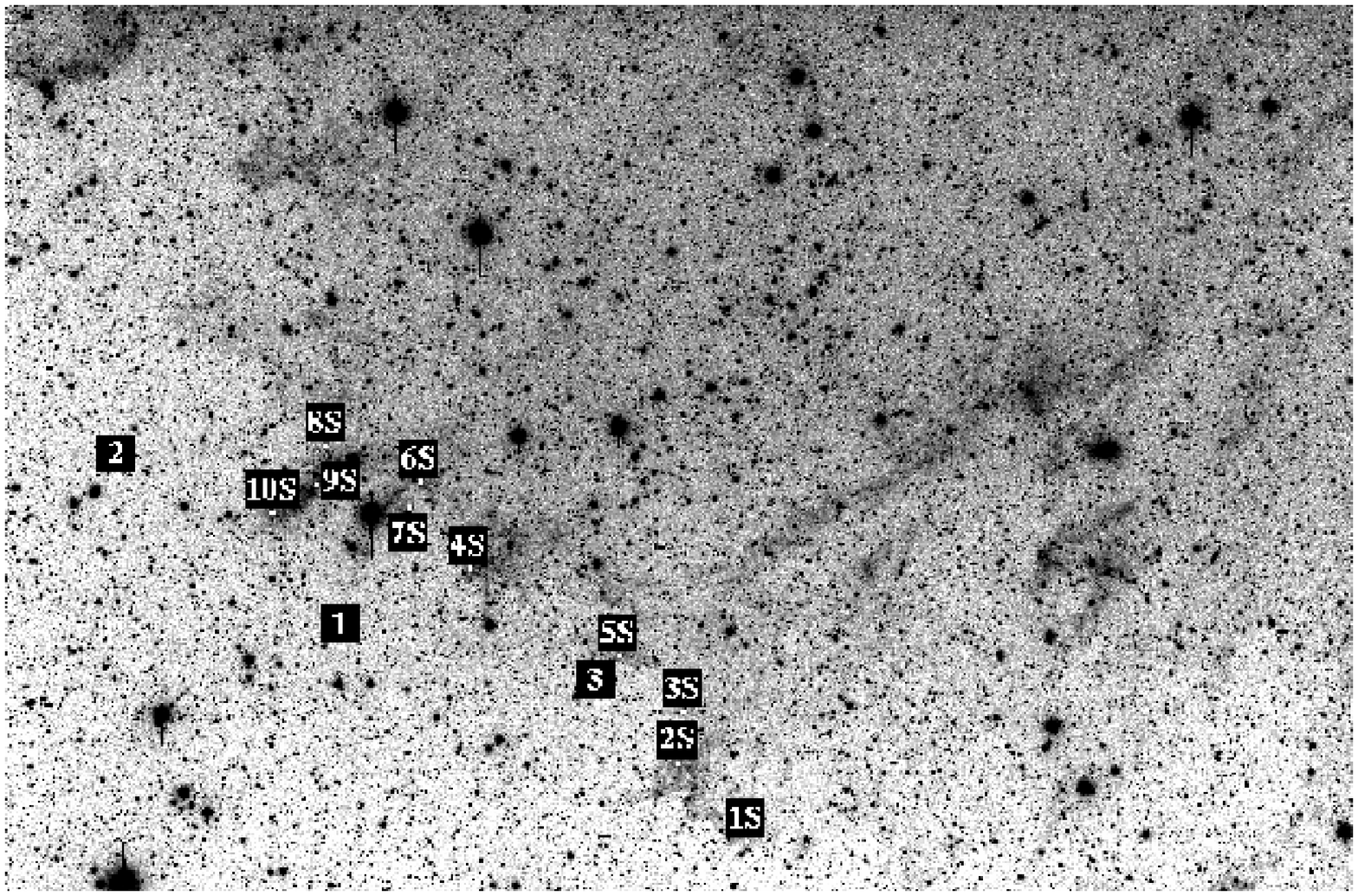}}

\begin{figure}
\centering
\includegraphics[width=6.5in]{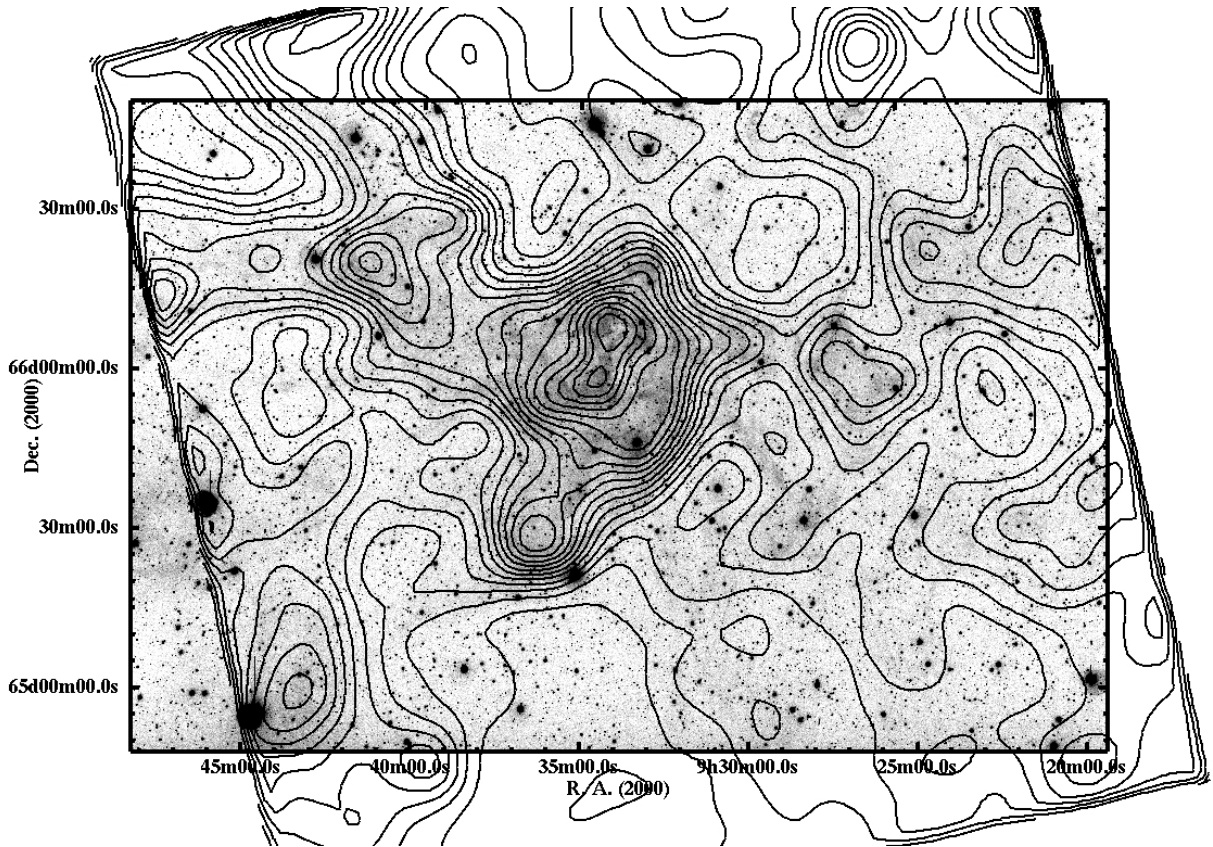}
\caption{The IRAS 100 \micron\ isophotes of MBM 32 from the map of Schegel et al. (1998) are superimposed on a 5400 sec exposure of the same object in the G-band (533 nm). The resolution in the optical image is 24 times higher than in the far-IR map.}
\end{figure}

\begin{figure}
\centering
\includegraphics[angle=-90,width=6.5in]{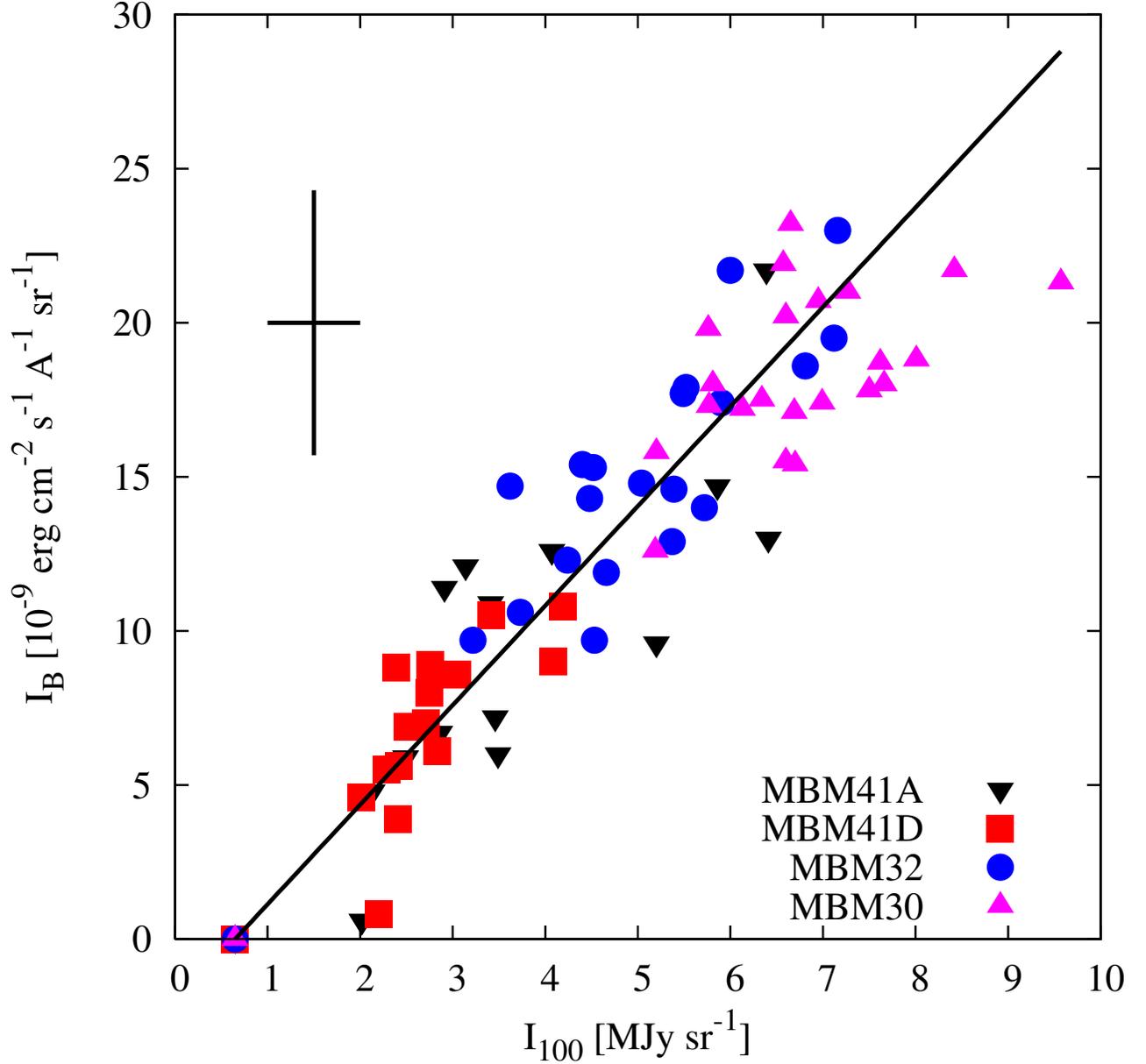}
\caption{The correlation between the B-band sky-subtracted nebular intensities and the IRAS 100 \micron\ intensities at identical positions from the map of Schlegel et al. (1998) for MBM 30, MBM 32, MBM41A, and MBM 41D.. The IRAS 100 \micron\ intensity at the sky reference level was 0.65 MJy sr$^{-1}$. The optical measurements represent averages over a square sampling aperture of 9 x 9 pixels. The error bar is one standard deviation of the mean of 81 pixels, and is strongly dominated by the noise in the foreground sky. The linear correlation indicates that the four clouds are optically thin in the B-band.}
\end{figure}

\begin{figure}
\centering
\includegraphics[width=6.5in]{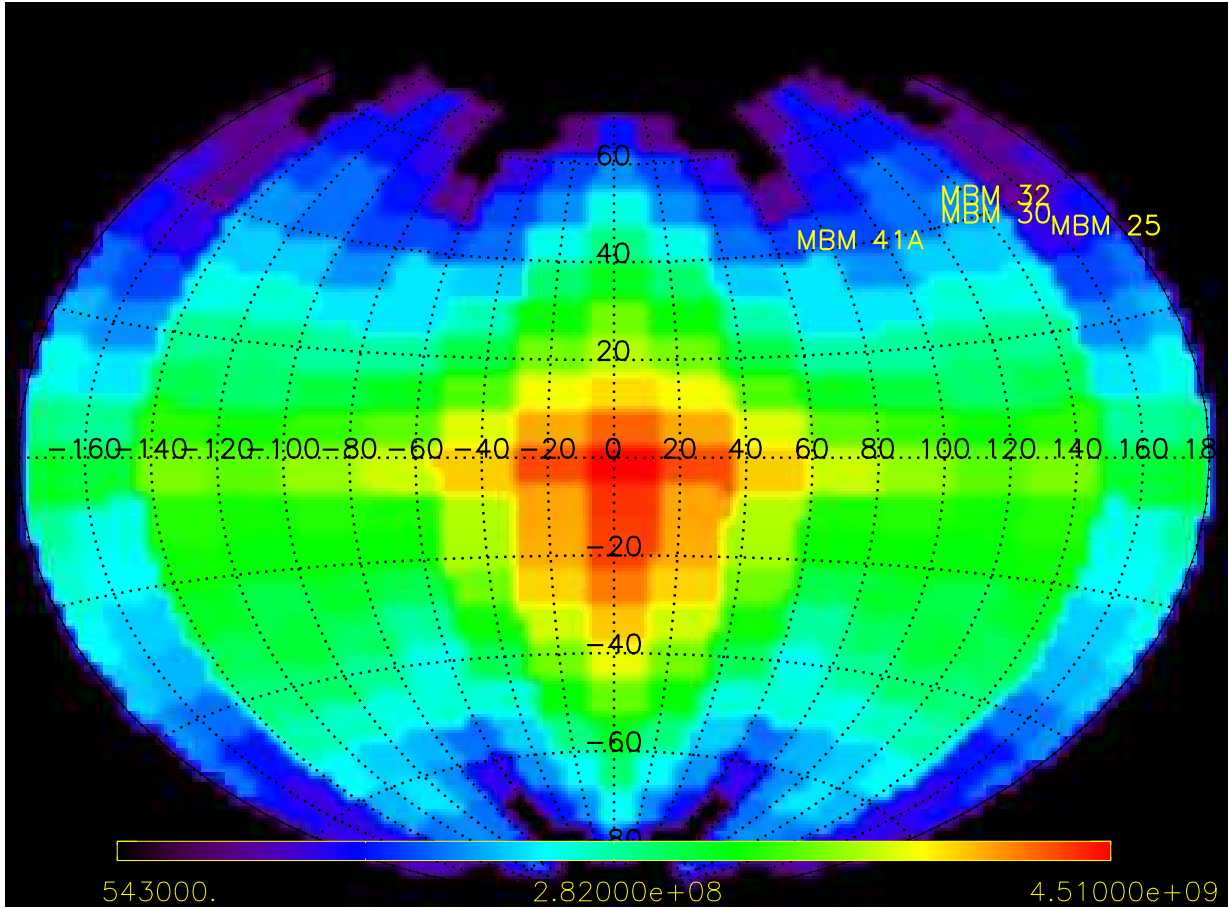}
\caption{Aithoff projection of the interstellar radiation field model of Porter \& Strong(2005) for the solar galactocentric radius and z = 100 pc for the B-band. The projection is centered on the direction toward the Galactic center and the approximate positions of our HGLICs are indicated. Galactic longitudes increase toward the right.}
\end{figure}

\begin{figure}
\centering
\includegraphics[width=6.5in,clip]{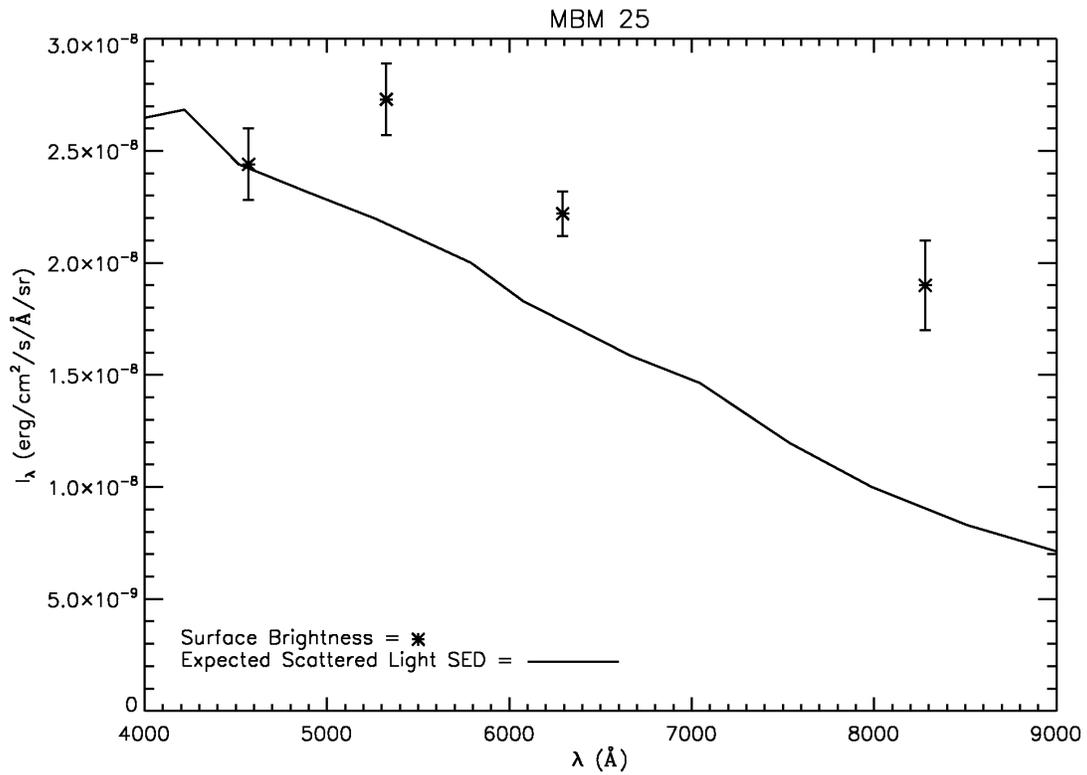}
\caption{(a - e) The sky-corrected average surface brightness intensities of MBM25, 30, 32, 41A, and 41D are compared with the expected scattered light SEDs computed for the respective optical depths of the clouds. Note that the error bars are smaller than in Fig. 4. Fig. 4 shows individual measurements of averages over 81 pixels while these figures display averages over ten measurements on individual ares containing 49 pixels.}
\end{figure}
\centering{
\includegraphics[width=6.5in,clip]{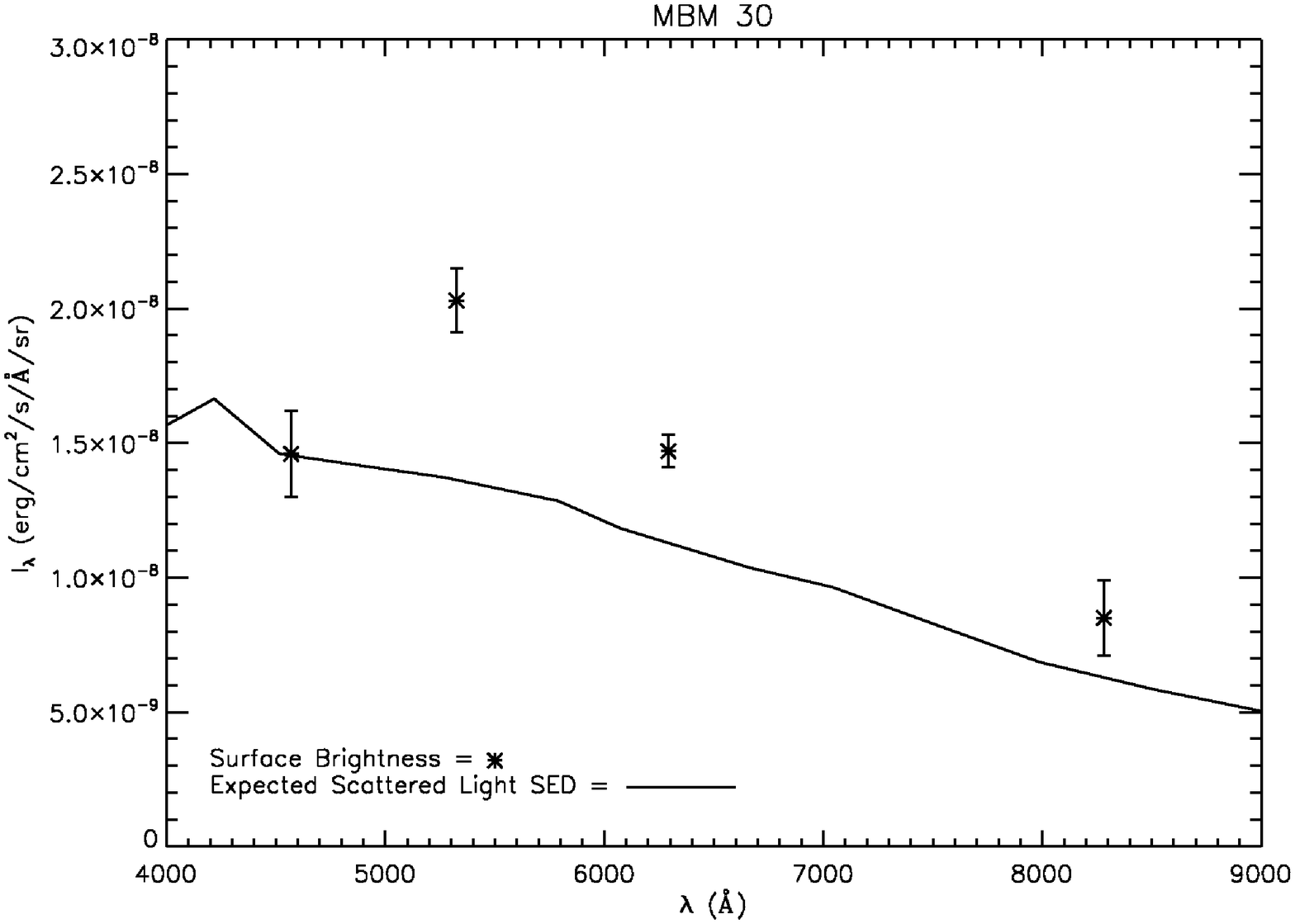}
\includegraphics[width=6.5in,clip]{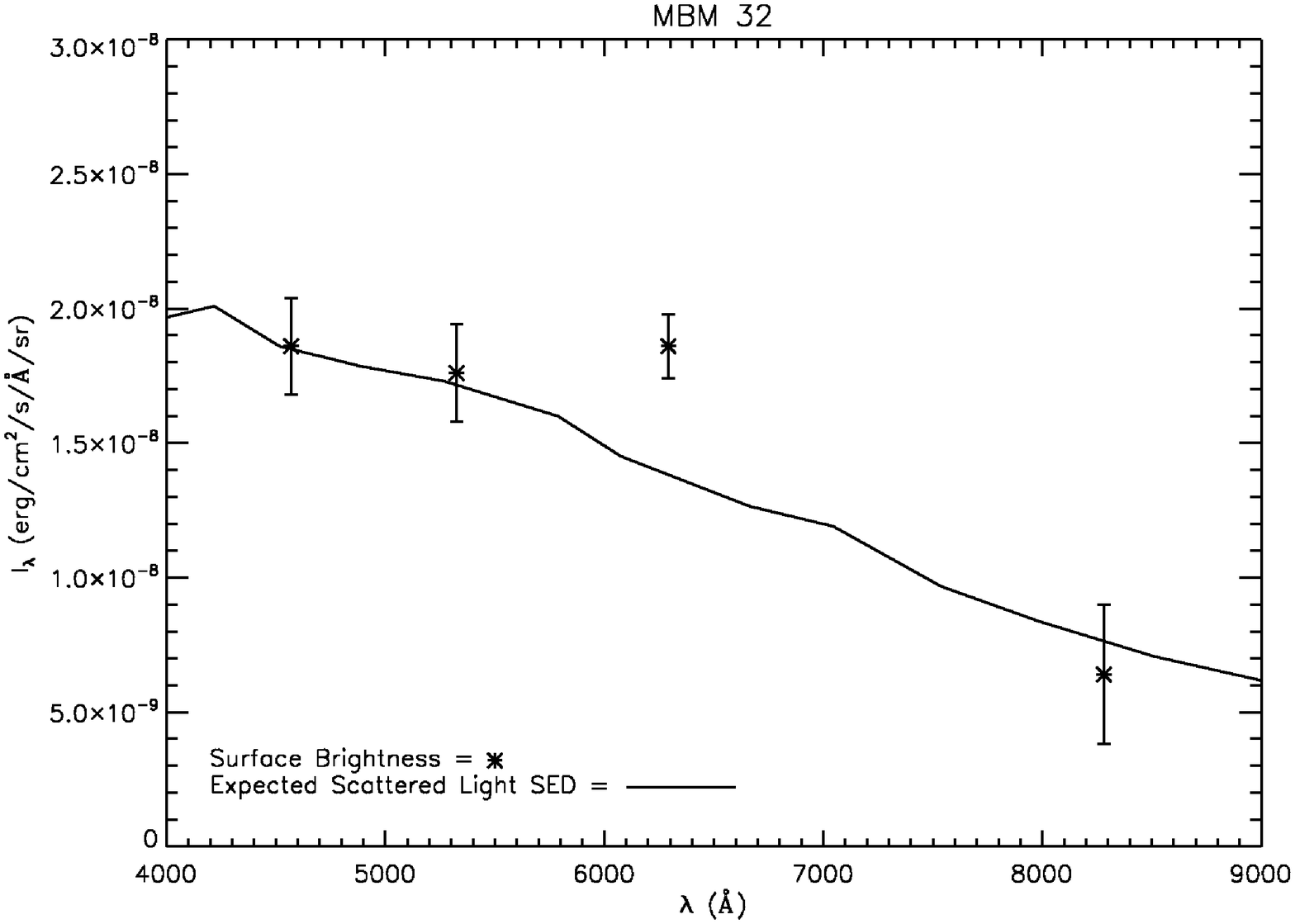}
\includegraphics[width=6.5in,clip]{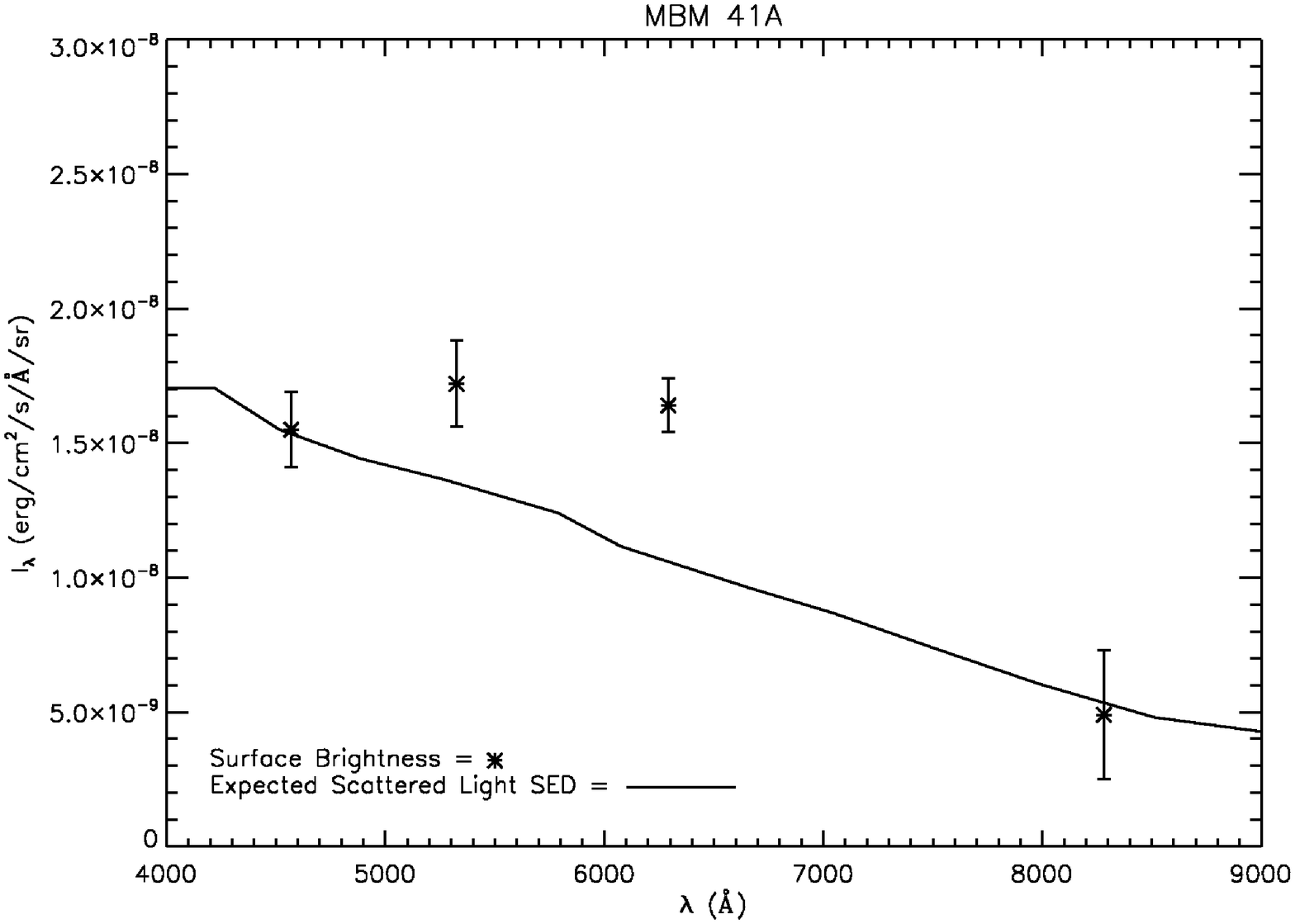}
\includegraphics[width=6.5in,clip]{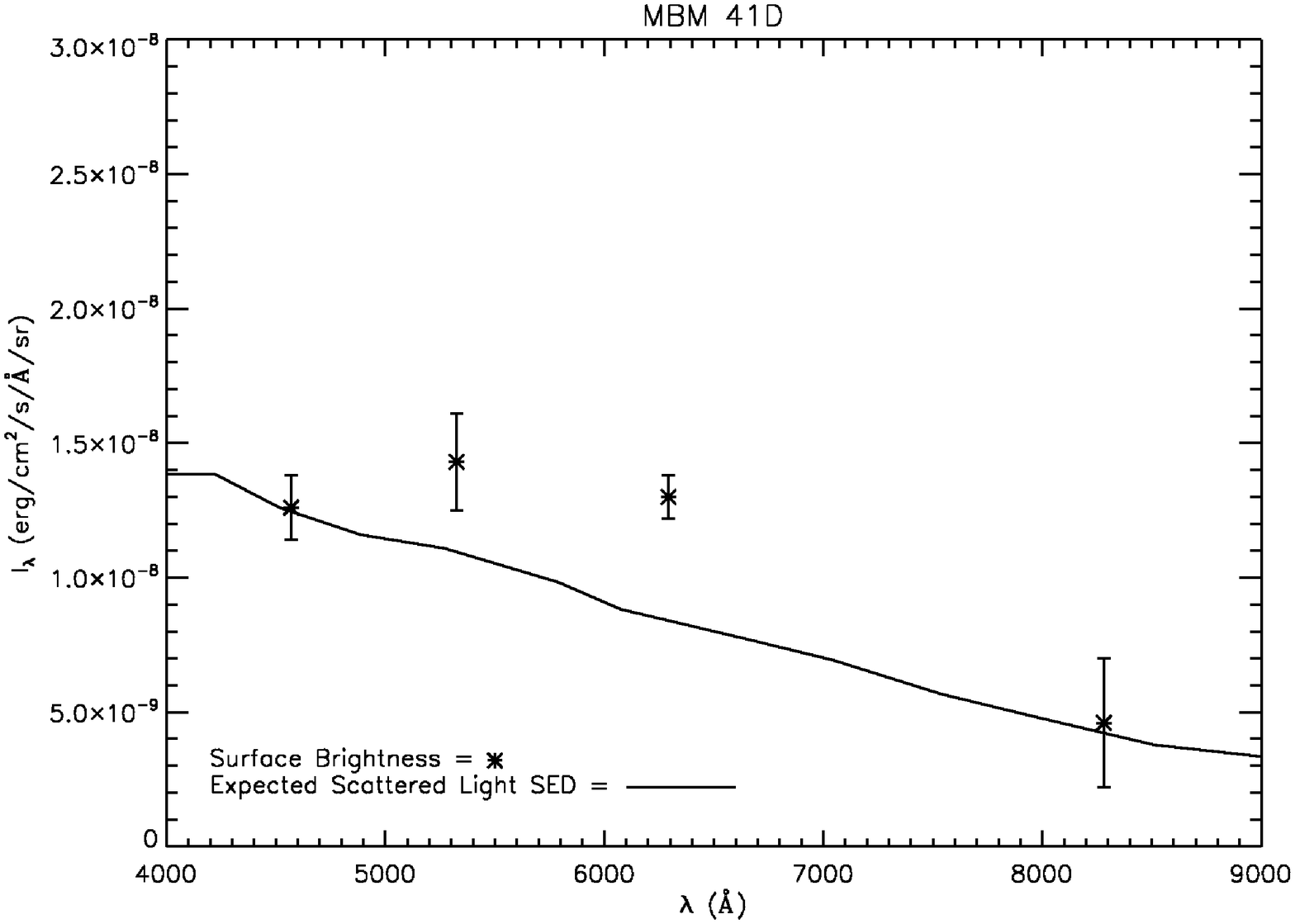}}

\clearpage

\begin{deluxetable}{ccc}
\tablecaption{Filter Characteristics\label{tbl-1}}
\tablewidth{0pt}
\tablehead{
\colhead{Filter} & \colhead{$\mathrm{\lambda_{eff}\ (\AA)}$} & \colhead{FWHM ($\mathrm{\AA}$)}
}
\startdata
B	& 	4570		&	1218\\
G	&	5325		&	714\\
R	&	6293		&	1244\\
I	&	8281		&	860\\
H$_\alpha$ &	6563	&	60\\
\enddata
\end{deluxetable}

\clearpage

\begin{deluxetable}{cccc}
\tablecaption{Basic Data for Observed Objects\label{tbl-2}}
\tablewidth{0pt}
\tablehead{
\colhead{Object} & \colhead{Gal. long.} & \colhead{Gal. Lat.} & \colhead{V$_\mathrm{LSR}$ [km s$^{-1}$]}
}
\startdata
MBM 25	&	173.8		&		+31.5		&	-8.1\\
MBM 30	&	142.2		&		+38.2		&	+2.7\\
MBM 32	&	147.2		&		+40.7		&	+4.0\\
MBM 41A	&	 90.0			&		+39.0		&	-23.0\\
MBM 41D	&	 92.3			&		+37.5		&	-23.0\\
\enddata
\end{deluxetable}

\clearpage

\begin{deluxetable}{lccccc}
\tablecaption{Dates of Observations and Total Integration Times\label{tbl-3}}
\tablewidth{0pt}
\tablehead{
\colhead{Object} & \colhead{B} & \colhead{G} & \colhead{R} & \colhead{H$_\alpha$} & \colhead{I}
}
\startdata
MBM 25			&	1/6/2006	&	1/6/2006	&	1/7/2006	&	1/9/2006	&	1/7/2006\\
$\left[sec\right]$	&	7200		&	5400		&	7200		&	7200   	&       7200\\
\\
MBM 30			&	2/28/2006	&	3/5/2006	&	3/5/2006	&	2/7/2006     &	3/25/2006\\
$\left[sec\right]$	&	7200		&	7200		&	7200		&	10800          &    15300\\
\\
MBM 32			&	1/28/2006	&	1/28/2006	&	1/28/2006	&	1/29/2006    &       1/31/2006\\
$\left[sec\right]$	&	7200		&	5400		&	5400 	&	7200             &                8100\\
\\
MBM 41-A		&	3/29/2006	&	3/29/2006	&	3/30/2006	&	4/9/2006         &           4/21/2006\\
$\left[sec\right]$	&	7200		&	6300		&	7200		&	14400             &             10800\\
\\
MBM 41-D		&	4/28/2006	&	4/28/2006	&	4/28/2006	&	4/30/2006 	&          4/30/2006\\
$\left[sec\right]$	&	9000		&	7200		&	8100		&	7200           	&                  10800\\
\enddata
\end{deluxetable}

\clearpage

\begin{deluxetable}{ccccc}
\tablecaption{Averages of Measured Surface Brightnesses\label{tbl-4}}
\tablewidth{0pt}
\tablehead{
\colhead{Object} & \colhead{B} & \colhead{G} & \colhead{R} & \colhead{I}  \\
\colhead{} & \multicolumn{4}{c}{$\left[\mathrm{erg\ cm^{-2}\ s^{-1}\ \AA^{-1}\ sr^{-1}}\right]$}
}
\startdata
	MBM 25	&	26.7$\pm$1.6 $\times 10^{-9}$	&	27.5$\pm$1.6 $\times 10^{-9}$	&	22.3$\pm$1.0 $\times 10^{-9}$	&	19.0$\pm$2.0 $\times 10^{-9}$     \\
	MBM 30	&	14.5$\pm$1.6 $\times 10^{-9}$	&	20.3$\pm$1.2 $\times 10^{-9}$	&	14.3$\pm$0.6 $\times 10^{-9}$	&	8.5$\pm$1.4 $\times 10^{-9}$       \\     
	MBM 32	&	18.5$\pm$1.8 $\times 10^{-9}$	&	17.6$\pm$1.8 $\times 10^{-9}$	&	19.7$\pm$1.2 $\times 10^{-9}$	&	5.6$\pm$2.6 $\times 10^{-9}$       \\
	MBM 41A	&	16.4$\pm$1.4 $\times 10^{-9}$	&	18.0$\pm$1.6 $\times 10^{-9}$	&	17.4$\pm$1.0 $\times 10^{-9}$	&	5.4$\pm$2.4 $\times 10^{-9}$       \\		
	MBM 41D	&	12.6$\pm$1.2 $\times 10^{-9}$	&	14.3$\pm$1.8 $\times 10^{-9}$	&	13.0$\pm$0.8 $\times 10^{-9}$	&	4.6$\pm$2.4 $\times 10^{-9}$       \\
\enddata
\end{deluxetable}

\clearpage

\begin{deluxetable}{ccccccc}
\tablecaption{Adopted Optical Depths\label{tbl-5}}
\tablewidth{0pt}
\tablehead{
\colhead{Object} & \multicolumn{4}{c}{$I_{100}\ \micron \left[\mathrm{MJy\ sr}^{-1}\right] $}  & \multicolumn{2}{c}{$\tau_B$} \\
\colhead{} & \colhead{cloud} & \colhead{sky} & \colhead{cloud-sky} & \colhead{cloud/sky} & \colhead{cloud} & \colhead{sky} 
}
\startdata
	MBM 25	&	 3.67          &        1.94       &       1.73           &         1.89		&	0.44	&	0.23\\
	MBM 30	&	 6.29          &        2.25       &       4.04           &         2.80		&	0.75	&	0.27\\     
	MBM 32	&	 5.99          &        2.47       &       3.52           &         2.43		&	0.71	&	0.29\\
	MBM 41A	&	 4.10          &         0.92      &       3.18           &         4.46		&	0.49	&	0.11\\		
	MBM 41D	&	 2.75          &         1.30      &       1.45           &         2.12		&	0.33	&	0.15\\
\enddata
\end{deluxetable}

\clearpage

\begin{deluxetable}{cccc}
\tablecaption{Input Parameters for Cloud SEDs\label{tbl-6}}
\tablewidth{0pt}
\tablehead{
\colhead{$\lambda$ (nm)} & \colhead{$J_\lambda\ (\mathrm{erg\ cm^{-2}\ s^{-1}\ \AA^{-1}\ sr^{-1}})$} & \colhead{$\tau_\lambda/\tau_{454}$} & \colhead{Albedo$_\lambda$}
}
\startdata
337	&		1.32 (-7)		&		1.29		&		0.670\\
390	&		1.44 (-7)		&		1.14		&		0.670\\
420	&		1.50 (-7)		&		1.08		&		0.670\\
454	&		1.42 (-7)		&		1.00		&		0.670\\
493	&		1.45 (-7)		&		0.90		&		0.657\\
527	&		1.50 (-7)		&		0.82		&		0.643\\
579	&		1.53 (-7)		&		0.73		&		0.623\\
607	&		1.51 (-7)		&		0.69		&		0.590\\
666	&		1.50 (-7)		&		0.62		&		0.563\\
705	&		1.48 (-7)		&		0.57		&		0.549\\
755	&		1.40 (-7)		&		0.53		&		0.516\\
802	&		1.34 (-7)		&		0.47     	&   		0.496\\
848	&		1.28 (-7)		&		0.42		&		0.469\\
918	&		1.22 (-7)		&		0.36		&		0.449\\
973	&		1.15 (-7)		&		0.33		&		0.422\\
\enddata
\end{deluxetable}

\end{document}